%% file: Arxiv_Submission.tex
\documentclass[aps,pre,twocolumn,superscriptaddress,groupedaddress]{revtex4-1} 

\usepackage{graphicx}  
\usepackage{dcolumn}   
\usepackage{bm}        
\usepackage{bbold}
\usepackage{appendix}
\usepackage{amssymb, amsmath}   
\usepackage[usenames, dvipsnames]{color}
\usepackage{float}

\begin{document}

\title{Chaotic dynamics in a single excitation subspace: deviations from the ETH via long time correlations}
\author{Charlie Nation}%
 \email{C.Nation@sussex.ac.uk}
 \affiliation{Department of Physics and Astronomy, University College London, London WC1E 6BT, United Kingdom}
\affiliation{%
	Department of Physics and Astronomy, University of Sussex, Brighton, BN1 9QH, United Kingdom.
}%
\author{Diego Porras}%
 \email{D.Porras@iff.csic.es}
\affiliation{%
 Institute of Fundamental Physics, IFF-CSIC, Calle Serrano 113b, 28006 Madrid, Spain 
} 
\date{\today}

\begin{abstract}
In this work we study a scenario where dynamics is restricted to a single excitation manifold, for particular physical observables with support in the manifold, which we label a `correlated quench'. We ask how such dynamics may in general differ from predictions of the eigenstate thermalization hypothesis (ETH).
We show that if thermalization occurs, it will not fulfil other key predictions of the ETH; instead following differing generic behaviours.
We show this by analysing long-time fluctuations, two-point correlation functions, and the out-of-time-ordered correlator; analytically detailing deviation from ETH predictions. We derive instead an ETH-like relation, with non-random off-diagonals, for matrix elements of observables, with correlations that alter long-time behaviour and constrain dynamics.  
Further, we analytically compute the time-dependence of the decay to equilibrium, showing that it is proportional to the survival probability of the initial state. 
We finally note that the conditions studied are common in many physical scenarios, such as under the rotating-wave approximation. We show numerically that predictions are robust to perturbations that break this approximation.
\end{abstract}

\maketitle

\section{Introduction}

The thermalization of closed quantum systems has seen a large amount of interest in recent years \cite{Rigol2008,Yukalov2011, Eisert2015,DAlessio2016, Gogolin2016, Borgonovi2016}, inspired by the modern experimental capability to observe their unitary evolution in the laboratory
\cite{Gring2012,Schneider2012,Georgescu2014, Schreiber2015,  Neill2016,  Kaufman2016a, Clos2016a}. 
For non-integrable systems, the Eigenstate Thermalization Hypothesis (ETH) has become the leading contender for a physical mechanism for thermalization \cite{Deutsch1991, Srednicki1996, Rigol2008, Kim2014, DAlessio2016, Mondaini2016, Mondaini2017}. The ETH can be written as a condition for the matrix elements of observables of a quantum many-body system. 
This conjecture can be motivated from random matrix theory by assuming that a non-integrable quantum Hamiltonian can be expressed as a perturbation of a non-interacting or integrable model. As shown by J.M. Deutsch in \cite{Deutsch1991}, this random matrix approach can be used to prove that time-averages of a typical observable are equivalent to microcanonical averages, which is one of the conditions required for a quantum system to thermalize. 
The second requirement for quantum thermalization is that observable fluctuations decrease exponentially with system size. This is also guaranteed by the ETH in the form of Srednicki's ansatz \cite{Srednicki1996, Srednicki1999}.

The ETH implies that generic closed chaotic quantum systems display many universal behaviours, independent of any particular form of the Hamiltonian. These universal behaviours are closely linked with the behaviours of random matrix models \cite{Deutsch1991, Reimann2008, Nation2018, Nation2019, Dabelow2019}. One may expect, however, that the presence of conservation laws may cause deviations from quantum chaos as described by RMT, with the most obvious example being integrable systems, with an extensive number of conserved quantities \cite{Cassidy2011}; what is not obvious is the possible effect of one, or a few, conservation laws on the route to equilibrium and other markers of thermalization.

We will see that for given realistic initial states and observables a single conserved quantity has a profound effect on dynamics, causing a departure from the expected behaviours implied by the ETH, {\it even in the case of non-integrable systems}. We study this departure in three key ways: i) fluctuations from equilibrium, ii) behaviour of two-time correlation functions, iii) scrambling of quantum information. We further study the time-dependence of observables after a quantum quench - a topic which has seen significant recent interest \cite{Malabarba2014, Garcia-Pintos2017, Reimann2016, Nation2019, Borgonovi2019, Dabelow2019, Alhambra2019}. We obtain an explicit form for the time-dependence in terms of the survival probability of the initial state. 

We stress that our claim is not that the introduction of a symmetry in generic closed systems violates the ETH. In fact, the ETH can be seen to still hold in each symmetry sector. Our findings indicate that the symmetry introduces a set of initial states, that happen to be common in many physical systems, that themselves have behaviour that deviates from the ETH prediction in the three ways described above.

We further note that the OTOC has been previously understood as a witness of quantum phase transitions: in Ref. \cite{Dag2019} the long-time value of the OTOC was seen to probe quantum phase transitions in non-integrable quantum spin chain system. Here the authors observe that in one phase the OTOC is approximately described by its ground state contribution at long times, as contributions by excited states are suppressed. Indeed, in the following we observe a phenomenologically similar mechanism, where contributions to the long-time OTOC are suppressed by the choice of observable and initial state.

This article is arranged as follows. First, we review some generic properties of non-integrable systems implied by the ETH, that are to be studied later in the presence of conservation law. 
We then outline a very simple case where such generic properties can be seen to differ from the ETH prediction, giving an intuition for the main physical mechanism behind the deviation. 
We then apply this approach to more general observables, and show that it leads to a scaling of time fluctuations that violates the ETH prediction. 
We then use the same approach developed for long-time fluctuations to understand both the equilibration in time of observables, and the scrambling of quantum information. 
Throughout, we present exact diagonalization calculations to demonstrate our analytical arguments.

\section{Properties of Chaotic Quantum Systems}
In this section we review some generic properties of non-integrable systems that are assumed to abide by the ETH. A typical scenario, that we will consider in this work, is that of a `quantum quench'. In this case, we start with some non-interacting Hamiltonian $H_0$, and prepare the system in and eigenstate $|\phi_{\alpha_0}\rangle$ of $H_0$. At $t = 0$, an interaction Hamiltonian $V$ is introduced, which renders the total Hamiltonian $H = H_0 + V$ non-integrable.
The many-body Hamiltonian, $H$, has eigenstates and eigenvalues $|\psi_\mu\rangle$  and $E_\mu$, respectively. 

To simplify the discussion, we focus on observables $O$ that have a diagonal structure in the basis of $H_0$. Initialized in a state $|\psi(0)\rangle = |\phi_{\alpha_0}\rangle = \sum_\mu c_\mu(\alpha_0) |\psi_\mu\rangle$, the initial observable expectation value $\langle O(t) \rangle = O_{\alpha_0 \alpha_0}$. After the perturbation $V$ is turned on, the system thermalizes; that is, observables evolve to an equilibrium given by the microcanonical ensemble $\langle O \rangle_{MC} = \langle O \rangle_{MC}(E_{\alpha_0})$, which depends only on the initial state energy.

The ETH is a conjecture on the properties of chaotic systems that provides a mechanism for thermalization, and can be written as an ansatz on observable matrix elements, 
$O_{\mu\nu} := \langle\psi_\mu|O|\psi_\nu\rangle$ in the energy eigenbasis \cite{Srednicki1996, Srednicki1999}:
\begin{equation}\label{eq:ETH}
O_{\mu\nu} = {\cal O}(E)\delta_{\mu\nu} + \frac{1}{\sqrt{D(E)}}f(E, \omega){\cal R}_{\mu\nu},
\end{equation}
where $E = \frac{E_\mu + E_\nu}{2}$, $\omega = E_\mu - E_\nu$, $f$ and ${\cal O} \approx \langle O \rangle_{MC}$ are smooth functions of their respective parameters. The function $f(E,\omega)$ has a energy width that determines the energy window under which two energy eigenstates have a non-negligible observable matrix element.
$D(E)$ is the density of states at energy $E$, and ${\cal R}_{\mu\nu}$ is a stochastic variable of zero mean and unit variance. 

Chaotic quantum systems are often characterized by an effective description in terms of random matrix theory (RMT) \cite{Deutsch1991, Flambaum1997, Reimann2015, Borgonovi2016, Torres-Herrera2016, Nickelsen2019}. Indeed, it was recently shown by us that the full form of Eq. \eqref{eq:ETH} can be derived in full from RMT methods \cite{Nation2018}, as can the full decay process, and subsequent fluctuations \cite{Nation2019, Nation2020}. Properties of random matrix models thus provide a powerful heuristic tool, from which we may understand the properties of non-integrable systems analytically. In the remainder of this section, we will describe three key features of such systems, that may be understood from the ETH and RMT.

\subsection{Thermalization}

\subsubsection{Equilibration to a thermal state}
The first key feature of chaotic systems is the tendency in time to an equilibrium state described by a thermal ensemble. This has motivated the study of chaotic quantum systems as a fundamental description of the emergence of statistical physics from many-body dynamics \cite{Neumann2010, Goldstein2009, Eisert2015}. The ETH describes sufficient conditions for thermalization of the long-time average value of an observable, which can be shown for an arbitrary initial state $|\psi(0)\rangle = \sum_\mu c_\mu |\psi_\mu\rangle$ with energy $E_0$ as follows:
\begin{align}
\overline{\langle O(t) \rangle } &:= \lim_{T\to \infty} \frac{1}{T}\int_0^T dt \langle O(t)\rangle \nonumber \\&
= \lim_{T\to \infty} \frac{1}{T}\int_0^T dt \sum_{\mu, \nu} c_\mu c^*_\nu O_{\mu, \nu} e^{-i(E_\mu - E_\nu)t} \\ &
= \sum_\mu |c_\mu|^2 O_{\mu, \nu} \nonumber \\&
\approx {\cal O}(E_0), \nonumber
\end{align}
where in the third line we have assumed that there are no degenerate eigenenergies, and in the last line assumed the ETH, and that the initial state is not a macroscopic superposition, such that the energy variance is small compared to macroscopic energies.

\subsubsection{Long-Time Fluctuations}

A second condition for thermalization to occur is the exponential vanishing of long-time fluctuations with system size. That is, after equilibration to a thermal state, fluctuations around this state in time should be small. We define the long-time fluctuations of an observable $O$ as
\begin{equation}
 \delta_O^2(\infty) := \lim_{T\to \infty} \left[\frac{1}{T} \int_0^t dt \langle O(t) \rangle^2 - \left(\frac{1}{T}\int_0^t dt \langle O(t) \rangle\right)^2\right].
\end{equation}
Assuming that there are not an extensive number of degeneracies, we use the the diagonal ensemble (DE) result \cite{DAlessio2016, Cassidy2011, Vidmar2016a},
\begin{equation}\label{eq:DE_flucs}
\delta_O^2(\infty) = \sum_{\substack{\mu\nu \\ \mu\neq \nu}} |c_\mu(\alpha_0)|^2|c_\nu(\alpha_0)|^2|O_{\mu\nu}|^2,
\end{equation}
where the index $\alpha_0$ indicates the initial state, $|\phi_{\alpha_0}\rangle = \sum_\mu c_\mu(\alpha_0)|\psi_\mu\rangle$, and $O_{\mu\nu} = \langle\psi_\mu|O|\psi_\nu\rangle$ are the matrix elements of some observable $O$ in the eigenbasis $\{|\psi_\mu\rangle\}$ of the many-body Hamiltonian, $H$.

Indeed, the size of fluctuations at long-times has been well studied in the ETH regime \cite{Srednicki1999, DAlessio2016, Nation2018, Nation2019, Nation2019a, Borgonovi2017}, and in early works Reimann \cite{Reimann2008} and Short \cite{Short2011} provided bounds for the fluctuations in terms of some effective dimension of the state of the system. Here we use a similar effective system size, the Inverse Participation Ratio (IPR) \footnote{Note that our definition of IPR differs here from that in \cite{Clos2016a} and in other works in the field of quantum chaos, where
the reciprocal quantity is defined as the IPR. Our definition here is more consistent with the original notion of participation ratio as
the number of energy eigenstates or atomic orbitals involved in the initial state (see e.g. D J Thouless 1974 Phys. Rep.(section C of
Physics Letters) 13 93142).}, defined by
\begin{equation}
\textrm{IPR}(|\psi(0)\rangle) = \sum_\mu |\langle \psi_\mu|\psi(0)\rangle|^4,
\end{equation}
which can be seen to have reasonable properties, as for a totally localized state, with $c_\mu(\alpha) = \delta_{\mu\alpha}$, we have, $\textrm{IPR}(|\phi_\alpha\rangle) = 1$, and for a maximally delocalized state, with $c_\mu(\alpha) = \frac{1}{\sqrt{N}}$, we have  $\textrm{IPR}(|\phi_\alpha\rangle) = \frac{1}{N}$. The (inverse of) the IPR is also often referred to as the `number of principle components' \cite{Borgonovi2017, Borgonovi2019, Borgonovi2019a} for this reason.

Recently \cite{Nation2018}, the current authors have obtained a relationship between the DE fluctuations and IPR from a RMT approach, finding that for observables that are diagonal in the basis of eigenstates of the non-interacting part of a chaotic Hamiltonian (which are our focus in the current work),
\begin{equation}\label{eq:delta_IPR}
\delta^2_O(\infty) \propto \textrm{IPR}(|\psi(0)\rangle).
\end{equation}
Indeed, both Reimann and Short's bounds can be understood in terms of the IPR, and Eq. \eqref{eq:delta_IPR} can be seen to follow the same scaling implied by a saturation of their bounds. This scaling of the fluctuations with system size has also been argued as a direct consequence of the ETH in (the supplemental material of) Ref. \cite{Clos2016a}. Further, this relation links the vanishing of long-time fluctuations with notions of ergodicity in terms of the explored Hilbert space dimension \cite{Pietracaprina2017}, which may be measured by the IPR. 

\subsection{Correlations}

Here we discuss additional features of correlation functions that follow from the ETH, focussing on two-time correlation functions, and the OTOC. Previous results have 
 been derived for thermal or microcanonical averages \cite{Alhambra2019, Venuti2019}. To make the link to the current work, where we focus on initial pure states, we note results of `typicality' \cite{Goldstein2006, Reimann2007, Popescu2006} imply that `the overwhelming majority' of such initial pure states are representative of a relevant thermal ensemble \cite{Popescu2006}. We will see this numerically below, where we contrast our results for the correlated quench scenario to other physically relevant initial conditions.

\subsubsection{Two-time correlations}

The behaviour of two-time correlation functions, $\langle O(t) O\rangle$, is of fundamental and practical importance to the study of many-body systems, providing a foundation for linear response theory \cite{Kubo1966}, open quantum systems \cite{Breuer2002}, and many other approaches. The factorization of two-time correlation functions, 
\begin{equation}\label{eq:factorising}
\langle W(t\to \infty )V\rangle_{ens} = \langle W\rangle_{ens} \langle V \rangle_{ens} 
\end{equation}
where $\langle \cdots \rangle_{ens}$ denotes an average with respect to a relevant ensemble, is a fundamental result describing dissipative processes in quantum and classical systems. The factorization of two-time correlations has been derived from the ETH for both microcanonical \cite{Venuti2019} and thermal \cite{Alhambra2019} states. In the following we refer to initial states that result in a factorization of two-time correlators at long times as ergodic, thus extending previous formal definitions \cite{Venuti2019} to individual quantum states.

\subsubsection{Scrambling of Quantum Information}

In classical mechanics, chaos is identified by the exponential divergence of trajectories at a rate given by the Lyapunov exponent. That is, take an initial state, and a slightly perturbed copy - under chaotic time evolution these states exponentially diverge.

Chaos may similarly be measured in a quantum system by the spreading of local information about a state over the entire system in time. This can be measured, for example, by the evolution of a local observable $W(t) = e^{iHt}We^{-iHt}$, and the effect of a local perturbation $V$, by the so-called 
out-of-time-ordered correlator (OTOC) \cite{Maldacena2016, Hashimoto2017, Yan2019, Borgonovi2019a, Swingle2018, Iyoda2017, Huang2017, Foini2019}, defined as,
\begin{equation}
F(t) = \langle W^\dagger (t)V^\dagger W(t)V\rangle.
\end{equation}
Information scrambling refers to the spreading of local information over the degrees of freedom of a system, and is related to the chaoticity of quantum systems \cite{Maldacena2016}. 
The OTOC can be seen to be related to the commutator
\begin{equation}
\begin{split}
C(t) & := \langle |[W(t), V]|^2 \rangle \\&
= 2\left(1 - \textrm{Re}[F(t)]\right),
\end{split}
\end{equation}
for unitary $W, V$. For non-unitary $W, V$, there are additional time-ordered correlator terms in the $C(t)$. As scrambling of quantum information occurs as local information is delocalized over the many-body system, two initially commuting observables, $[W(0), V] = 0$, should have a non-zero commutator at some later time in a scrambling system. It can thus be seen that the OTOC, $F(t)$, is related to the growth of the support of localized operators in time, and information scrambling will cause $F(t)$ to decay.

As alluded above, the OTOC can be related to chaos in an analogous manner to the definition of a Lyapunov exponent in classical systems. In a chaotic quantum system at short times, the OTOC is expected to take the form \cite{Swingle2018},
\begin{equation}\label{eq:Lyapunov}
F(t) \approx 1 - \epsilon e^{-\lambda_L t},
\end{equation}
where $\lambda_L$ is conjectured to be a quantum analogue of the Lyapunov exponent.
The link between quantum chaos as defined by the ETH, and the behaviour of the OTOC in Eq. \eqref{eq:Lyapunov} is not yet totally clear, however some important steps have been made in e.g. linking the ETH to known bounds on $\lambda_L$ \cite{Murthy2019}.

In the following we limit our attention to the long-time average, defined by $\overline{F} := \lim_{T\to\infty}\frac{1}{T}\int_0^T dt F(t)$, which is expected to be equal to zero for chaotic systems. This may be seen for example from relating chaotic dynamics to random unitaries \cite{Roberts2017}, and more directly as a consequence of the ETH \cite{Huang2017}. We will see that when both the ergodicity relation \eqref{eq:factorising}, and the scaling of fluctuations Eq. \eqref{eq:delta_IPR}, deviate from that expected from the ETH, we similarly see a deviation of the long-time average $\overline{F}$ from the expected value of zero for chaotic systems.

Generally, the expectation value $\langle \cdots \rangle$ is taken at some inverse temperature $\beta$, such that $\langle \cdots \rangle = tr(\rho \cdots)$, with $\rho_{\mu\nu} \sim e^{-\beta E_\mu}\delta_{\mu\nu}$. Here, however, we focus on initial pure states, and thus use $\langle \cdots \rangle = \langle \psi |\cdots |\psi \rangle$ for some state $|\psi\rangle$.
$\overline{F}$ may be then written as,
\begin{equation}\label{eq:F_ave}
\begin{split}
\overline{F} & = \overline{\langle\phi_{\alpha_0}|W^\dagger (t)V^\dagger W(t)V|\phi_{\alpha_0}\rangle}
\\& = \sum_{\mu\nu}c_\mu(\alpha_0)c_\nu(\alpha_0) \overline{\langle \psi_\mu|W^\dagger (t)V^\dagger W(t)V|\psi_\nu\rangle},
\end{split}
\end{equation}
for an initial state $|\psi(0)\rangle = |\phi_{\alpha_0}\rangle = \sum_\mu c_\mu(\alpha_0)|\psi_\mu\rangle$. 
We assume time reversal symmetry, such that the $c_\mu(\alpha)$s may be taken as real, though one can easily see our approach extends to the non time-reversal symmetric case.

We note that it has been previously observed that scrambling is somewhat state-dependent \cite{Iyoda2017, Dag2019, Dag2020}, even in non-integrable systems - and we may expect a similar behaviour for two-time correlation functions - where initial pure states may deviate from the ETH result. The following results can be understood in the context of the ETH understanding that it necessarily omits eigenstate correlations - this has been previously shown to have an impact in terms of spatial correlations \cite{Chan2019}; the following results can be seen as a consequence of energetic correlations between the initial state, Hamiltonian, and observable.

\section{Set up}

In this section we describe, and give examples of, a scenario in which we observe a discrepancy from the markers of chaoticity in quantum systems outlined above. We will show this as a simple example of where one expects the scaling of fluctuations to differ from Eq. \eqref{eq:delta_IPR}, and outline more formally conditions where we expect this behaviour generally.

\subsection{Simple Example}
\begin{figure}
    \includegraphics[width=0.45\textwidth]{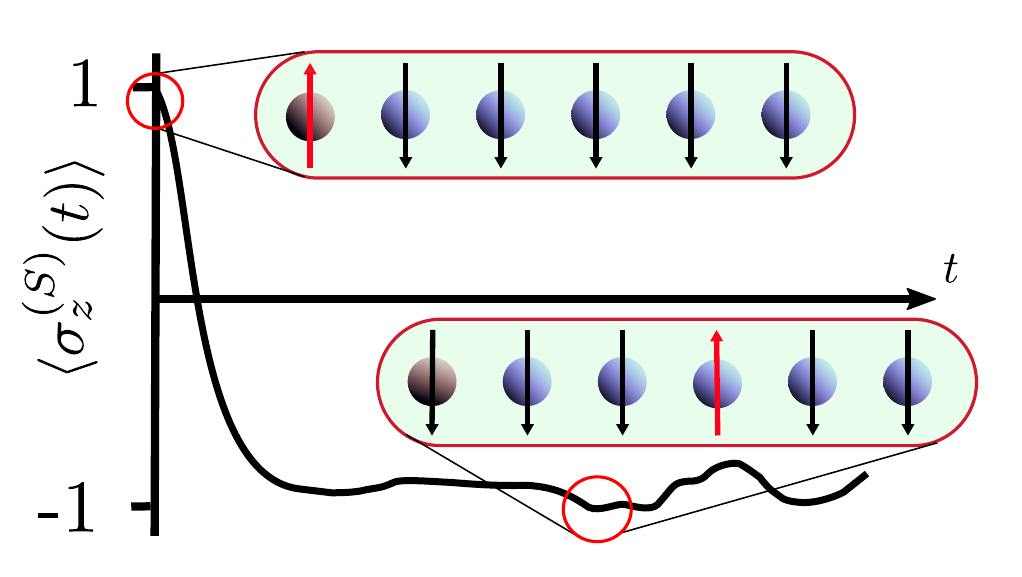}
    \caption{Idealised scenario of a correlated quench where $\hat{N} = \sum_i^N \sigma_z^{(i)}$.
It shows an initial state with a single spin excitation in the system and all spins in the bath in the ground state.  
If we assume that $\hat{N}$ is conserved, then at later times, the system spin is more likely to be in the ground state and a single excitation propagates in the bath. Note that the initial bath state is  not necessarily the ground state of $H_B$.
}
    \label{fig:drawing}
\end{figure}

We consider first a simple case of $N$ two level systems (qubits) with a Hamiltonian in the form $H = H_S + H_B + H_{SB}$, that conserves the total number of excited qubits, $\hat{N} = \sum_i^N \sigma_z^{(i)}$, where $\sigma_{\{x, y, z\}}^{(i)}$ are the Pauli matrices acting on site $i$. We initialize the system in the state $|\psi(0)\rangle = |\uparrow\rangle_S \prod_i^{{\cal N}_B} | \downarrow \rangle_{B, i}$. We thus see that the time evolution is restricted to the manifold of a single excitation only, as depicted in Figure \ref{fig:drawing}.

A following measurement of $\sigma_z^{(S)}$ is thus the same as a measurement of the survival probability $P_0$ of the initial state, as if the qubit is found to be excited, it is known that there are no excitations present in the bath, and vice versa. It can easily be seen that this does not extend to a local observable of the bath, and thus the relation between the local system observable and the global observable $P_0$ is due to correlations between the initial state and local observable. It is precisely these correlations that we are concerned with in this work.

The survival probability can be written as the expectation value of the operator $P_0$, where,
\begin{equation}
\begin{split}
P_0 &= |\psi(0)\rangle\langle\psi(0)| \\& 
 = \sum_{\mu, \nu} c_\mu(\alpha_0) c_\nu(\alpha_0) |\psi_\mu\rangle \langle \psi_\nu|, 
\end{split}
\end{equation}
where $|\psi_\mu\rangle$ is an eigenstate of $H$, and $|\psi(0)\rangle = |\phi_{\alpha_0}\rangle$, is the initial state, which may be written as $|\phi_{\alpha_0}\rangle = \sum_\mu c_\mu(\alpha_0) |\psi_\mu\rangle$. We thus have that $(P_0)_{\mu\nu} = c_\mu(\alpha) c_\nu(\alpha)$, such that the fluctuations, Eq. \eqref{eq:DE_flucs}, can be written as 
\begin{equation}\label{eq:NE_flucs_P}
\begin{split}
\delta_{P_0}^2(\infty) & = \sum_{\substack{\mu\nu\\ \mu\neq\nu}} |c_\mu(\alpha_0)|^2|c_\nu(\alpha_0)|^2 (c_\mu(\alpha_0)c_\nu(\alpha_0))^2\\&
= \sum_{\mu\nu}|c_\mu(\alpha_0)|^4|c_\nu(\alpha_0)|^4 - \sum_\mu |c_\mu(\alpha_0)|^8 \\&
\approx \mathrm{IPR}(|\psi(0)\rangle)^2,
\end{split}
\end{equation}
where in the last line we have used that many-body eigenstates of the systems of interest have a large number of principle components, and thus $\max c_\mu(\alpha) \ll 1$. 

One can thus see that for the survival probability, and thus for a class of observables related by a conservation law, we do not observe the scaling of fluctuations of Eq. \eqref{eq:delta_IPR}, predicted by RMT and the ETH. Indeed, it has recently been been shown that the survival probability is not `self-averaging' \cite{Schiulaz2019a}, such that it may not be expected to follow RMT behaviour at any time scale.

In this simple example lies the key intuition of the main result of this work, which we formulate in a more general scenario below. That is, for certain initial states and observables related to a conservation law, the long-time behaviour of the observable deviates from the expected behaviour due to the ETH. Here we have seen this for the example of observable fluctuations, however the more general treatment below will further analyse the behaviour of two- and four- point correlation functions. 

We once again stress that these results apply to systems that may in general obey the ETH. Indeed, in the example above $H$ may obey the ETH in each symmetry sector. Our result suggests that the introduction of a symmetry allows for correlations between the observable and initial state to dominate dynamics. It is these correlations that deviate the long-time behaviour from that expected by the ETH.

\subsection{Formal Conditions}

In the following, we focus on models that may be described by a Hamiltonian of the form $H = H_S + H_B + H_{SB}$,
where $H_S$ is a $2\times2$ Hamiltonian of a single qubit, with eigenstates $\{|\uparrow\rangle_S, |\downarrow\rangle_S\}$. We discuss in Appendix \ref{App:Larger_System} how these ideas scale to larger system Hamiltonians, noting that this is non-trivial, but can be expected in some generic settings. In the following we use subscripts $S$ and $B$ to refer to the system and bath respectively The subscript $SB$ denotes coupling terms between these subspaces.

One can see that a key condition of the discussion in the previous section is the presence of some conservation law $[H, \hat{N}] = 0$. We further require that the operator $\hat{N} = \hat{N}_S + \hat{N}_B$ is the sum of at least two local operators defined separately on the system and bath. These local operators are each conserved in both the system and bath under the non-interacting Hamiltonian $H_S + H_B$, such that $[H_S, \hat{N}_S] = 0$, and $[H_B, \hat{N}_B] = 0$. The `number of excitations' $N_{ex} = N_{S} + N_{B}$, for some state $|\psi\rangle$, is then $\langle \psi |\hat{N} |\psi\rangle = N_{ex}$. This quantity is conserved during time evolution, though $N_S$ and $N_B$ are not conserved individually by the coupling term $H_{SB}$. We study the behaviour of a local observable $O_S$ that is diagonal in the basis of the local excitation number, such that $[O_S,  \hat{N}_S] = [O_S, H_S] = 0$. 

The second key condition is that of the initial state. We require that the initial bath state is a non-degenerate state of the quantity $\hat{N}_B$, in the sense that there is a single state with the eigenvalue ${}_B\langle \psi(0)|\hat{N}_B|\psi(0)\rangle_B = N_B$. The system is initialized in the excited state of $\hat{N}_S$ (and thus of $H_S$). This ensures that, after measurement of $O_S$, if the system is found to be in the initial state, there is only a single excitation configuration possible for the bath state. This is guaranteed by choosing an initial state where there is a single excitation, localized to the system qubit.

To summarize, then, we focus on the behaviour of generic non-integrable systems under the following conditions: i) A conserved charge or excitation number, ii) The local system observable is diagonal in the basis of the local excitation number, iii) The initial state has a single excitation localized to the system. We will see below that these conditions are enough to identify a local observable with the survival probability (up to some constant factor), and thus ensure that the behaviour of the observable violates that predicted by the ETH, yet still thermalizes for systems with a large effective dimension.

In the following, we will refer to the thermalization under the above conditions as a `correlated quench'. We note that these conditions are indeed particularly restrictive, we will see, however, such specificity is to be expected - conditions in non-integrable systems that violate the ETH should be exceptionally rare. Further, we can note that as more symmetries are included, these conditions are not so stringent, as many more initial states fulfil them. This approach thus also contributes to a generic understanding of aspects of the behaviour of integrable and near integrable models, where conserved quantities dominate the dynamics. We further note that these conditions are applicable to many cases of interest, such as the Spin-Boson model, which we discuss in Appendix \ref{App:RWA}, and are observed to be robust to perturbations away from strict fulfilment conditions.

\section{Fluctuations and observable elements}

Here we focus on the case where the system is a single qubit with Hamiltonian $H_S$ with eigenstates $\{|\uparrow\rangle_S, |\downarrow\rangle_S\}$, and treat a generic bath with some conserved quantity $\hat{N}_B$. 
Applying, then, the conditions outlined above, we initialize our system in the state,
\begin{equation}
|\psi(0)\rangle = |\uparrow\rangle_S|k_0\rangle_B,
\label{initial.state}
\end{equation}
where $|k_\alpha\rangle_B$ denote eigenstates of $\hat{N}_B$, and $|k_0\rangle_B$ specifies a particular non-degenerate eigenstate of the conserved quantity $\hat{N}_B$, such as that with zero excitations, $N_B = 0$. We thus have $\langle \hat{N} \rangle = 1$. 
Now, we have for an arbitrary eigenstate of the interacting Hamiltonian, $H$,
\begin{equation}
|\psi_\mu\rangle = \sum_{\alpha}c_\mu(\alpha)|\phi_\alpha\rangle,
\end{equation}
where $|\phi_\alpha\rangle = |s_\alpha\rangle_S|k_\alpha\rangle_B$, with $s_\alpha = \{\uparrow, \downarrow\}$, is an eigenstate of the conserved quantity $\hat{N}$. For example, if in a system of $N$ qubits $\sum_i \sigma_z^{(i)}$ is conserved, then we may have $|\phi_\alpha\rangle = |\uparrow, \downarrow, \cdots \rangle$, and the conserved quantity is the total number of qubits in the up and down states. One may then, for example, define the number of excitations as $\hat{N} = \sum_i \frac{1}{2} (\sigma_z^{(i)} + \mathbb{1})$, equal to the number of qubits in the $|\uparrow\rangle$ state.

We separate the sum over $\alpha$ into two parts, those labelling the states with the system in each state $|\uparrow\rangle_S$ and $|\downarrow\rangle_S$, such that each part is a sum over the bath states $k$ (writing $c_\mu(\alpha) := c_\mu(k_\alpha, s_\alpha)$),
\begin{equation}
|\psi_\mu\rangle = \sum_{k}c_\mu(k, \uparrow)|\uparrow\rangle_S|k\rangle_B + \sum_{k}c_\mu(k, \downarrow)|\downarrow\rangle_S|k\rangle_B,
\end{equation}
where we have dropped the subscript $\alpha$.
Now, for a local system observable $O = O_S \otimes \mathbb{1}_B$ we have, for example, that ${}_B\langle k|{}_S\langle \uparrow |O|\uparrow \rangle_S| j \rangle_B = O_{\uparrow \uparrow}\delta_{kj}$, where $O_{\uparrow\uparrow} := {}_S\langle\uparrow|O_S|\uparrow\rangle_S$, and $O_{\downarrow\downarrow} := {}_S\langle\downarrow|O_S|\downarrow\rangle_S$. Thus, the matrix elements $O_{\mu\nu}$ may be expressed as
\begin{equation}\label{eq:Omunu_full}
\begin{split}
O_{\mu\nu} & = \sum_{k}c_\mu(k, \uparrow)c_\nu(k, \uparrow)O_{\uparrow\uparrow} + \sum_{k}c_\mu(k, \downarrow)c_\nu(k, \downarrow)O_{\downarrow\downarrow} \\ &
+ \sum_{k}c_\mu(k, \uparrow)c_\nu(k, \downarrow)(O_{\uparrow\downarrow} + O_{\downarrow\uparrow}).
\end{split}
\end{equation}
We can further use that
\begin{equation}\label{eq:useful_relation}
\begin{split}
\sum_{k} & c_\mu(k, \downarrow)c_\nu(k, \downarrow) \\& := \sum_{k}\langle\psi_\mu|\Big( |\downarrow\rangle_S|k\rangle_B {}_B\langle k |{}_S\langle \downarrow|\Big) |\psi_\nu \rangle \\&
=\langle \psi_\mu |\Big( \mathbb{1} - \sum_{k} |\uparrow\rangle_S|k\rangle_B {}_B\langle k |{}_S\langle \uparrow| \Big)|\psi_\nu\rangle \\& 
= \delta_{\mu\nu} - \sum_{k}c_\mu(k, \uparrow)c_\nu(k, \uparrow),
\end{split}
\end{equation}
where we have used the completeness relation $\sum_\alpha |\phi_\alpha\rangle\langle\phi_\alpha| = \mathbb{1} $, to obtain, for the case of an observable that commutes with the local excitation number (such that $O_{\uparrow\downarrow} = 0$),
\begin{equation}\label{eq:O_full}
O_{\mu\nu} = \Delta O \sum_{k}c_\mu(k, \uparrow)c_\nu(k, \uparrow) + O_{\downarrow\downarrow}\delta_{\mu\nu},
\end{equation}
where we have defined $\Delta O := O_{\uparrow\uparrow} - O_{\downarrow\downarrow}$.

\begin{figure*}
    \includegraphics[width=\textwidth]{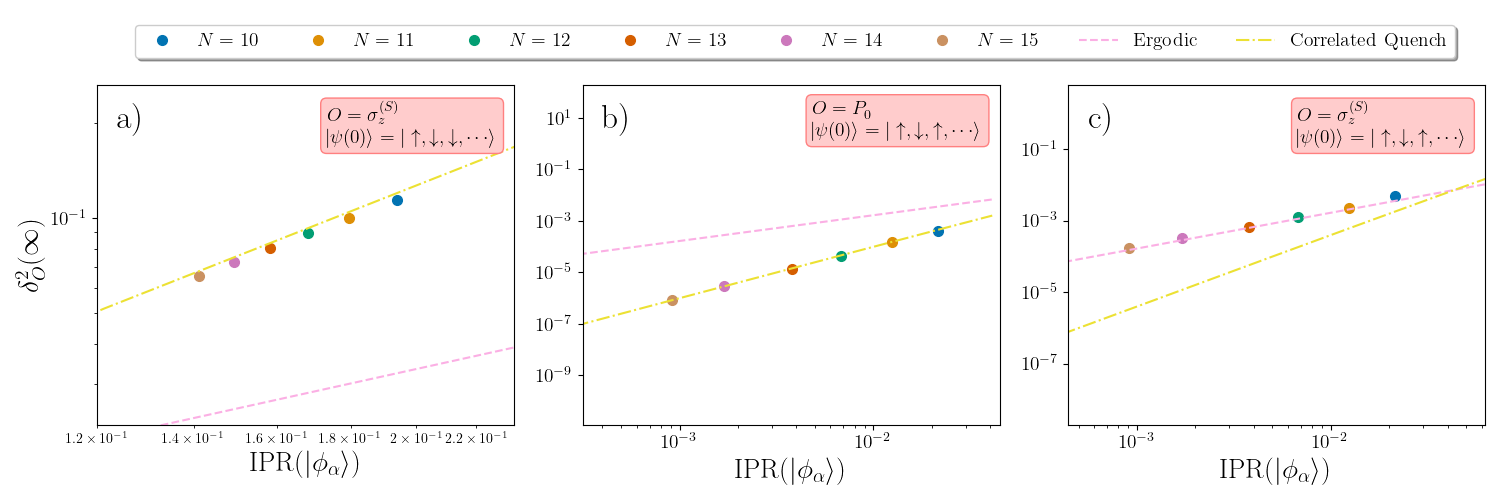}
    \caption{DE Fluctuations versus IPR for Hamiltonian \eqref{eq:XXX_H_0}, \eqref{eq:XXX_H_I}. $B_x = 0.05$, $J = 1$, $J^\prime = 0.8$. Observables are $\sigma_z^{(S)}$ for a), c) and $P_0$ for b). Initial bath states are all qubits down for a), and the Neel state for b), c).}
    \label{fig:XXX}
\end{figure*}

To gain an understanding of the effect of the conservation law, we can estimate the value of time-fluctuations assuming that the system is in the initial state 
\eqref{initial.state},
\begin{equation}
\begin{split}
\delta_{O}^2(\infty) &= 
\sum_{\substack{\mu,\nu \\ \mu \neq \nu}} |c_\mu(\uparrow,k_0)|^2 |c_\nu(\uparrow,k_0)|^2  \\&
 \times \Delta O^2 |\sum_k c_\mu(k,\uparrow)c_\nu(k,\uparrow)|^2.
\end{split}
\end{equation}
One could  naively assume that wave function components $c_\mu(k,\uparrow)$ and $c_\mu(k_0,\uparrow)$ are just independent random variables and carry out the summation. However, if we assume that the coupling term conserves the total number of excitations, $N_{ex}$, then the sum over $\mu$, $\nu$, must run over values with $N_{ex} = 1$. Thus, the components $c_\mu(k,\uparrow)$ can only take non-zero values if $k,\uparrow$ corresponds to a total excitation value $N_{ex} = 1$. Thus, the sum is restricted to $k = k_0$.

Concretely, then, we can see that the summation
\begin{equation}\label{eq:eta}
\eta = \sum_k c_\mu(k, \uparrow)c_\nu(k, \uparrow),
\end{equation}
may be restricted simply to the term,
\begin{equation}\label{eq:eta_ansatz}
\eta = c_\mu(k_0, \uparrow)c_\nu(k_0, \uparrow) := c_\mu(\alpha_0)c_\nu(\alpha_0),
\end{equation} 
where we have defined $\alpha_0 = (k_0, \uparrow)$ as the indices of the initial state.

This then leads us to the form for observable matrix elements,
\begin{equation}\label{eq:GSQ_ansatz}
O_{\mu\nu} \to \Delta O c_\mu(\alpha_0)c_\nu(\alpha_0) + O_{\downarrow\downarrow}\delta_{\mu\nu}.
\end{equation}
In-fact, one can see that this recovers the form of Srednicki's ansatz \cite{Srednicki1999} if $c_\mu(\alpha_0)c_\nu(\alpha_0)$ is taken to be a suitably small stochastic variable, however, Eq. \eqref{eq:GSQ_ansatz} allows for correlations between the wave function coefficients $c_\mu(k, \uparrow)$, to be included.

In understanding Eq. \eqref{eq:GSQ_ansatz} it is important to stress that the observable matrix elements are indeed still in reality described by Eq. \eqref{eq:O_full}, however, due to the conserved quantity $\hat{N}$ and correlated initial state the dynamics are restricted to a subset of the Hilbert space. It is precisely this restriction that allows us to make the substitution \eqref{eq:GSQ_ansatz}. 

Applying Eq. \eqref{eq:GSQ_ansatz} to the long-time fluctuations, we thus obtain,
\begin{equation}\label{eq:NE_flucs}
\begin{split}
\delta_O^2(\infty) & = \Delta O ^2 \sum_{\substack{\mu\nu\\ \mu\neq\nu}} |c_\mu(\alpha_0)|^2|c_\nu(\alpha_0)|^2 (c_\mu(\alpha_0)c_\nu(\alpha_0))^2\\&
= \Delta O ^2 \sum_{\mu\nu}|c_\mu(\alpha_0)|^4|c_\nu(\alpha_0)|^4 - \Delta O^2 \sum_\mu |c_\mu(\alpha_0)|^8 \\&
\approx \Delta O ^2\mathrm{IPR}(|\psi(0)\rangle)^2,
\end{split}
\end{equation}
where we have once more assumed that the eigenstates have a large number of principle components, and hence $\sum_\mu |c_\mu(\alpha_0)|^8 \ll (\sum_\mu |c_\mu(\alpha_0)|^4)^2$. We thus recover the same scaling of fluctuations as that seen in general for the survival probability, and differs from that obtained from the ETH.

We numerically confirm this approach in a quantum spin-chain model, described below, in Fig. \ref{fig:XXX}, where it is contrasted to the behaviour of a different initial state that is a highly degenerate eigenstate of $H_B$. We observe that the system observable scales as expected by \eqref{eq:delta_IPR} in this case, whereas the survival probability fluctuations deviate from the ETH for all initial states. In Appendix \ref{App:RWA} we study the Spin-Boson model, and see that the same scaling can be derived for this model in the rotating wave approximation.

Further, in Appendix \ref{App:robustness} we show a case where this scaling is obtained without the conservation of excitation number when the initial state is the ground state of a non-interacting Hamiltonian prior to a quantum quench. Intuitively, this can be seen as a by-product instead of conservation of energy; in cases where additional excitations require an additional energy cost, such transitions only contribute weakly, and Eq. \eqref{eq:GSQ_ansatz} holds approximately.

\section{Two-point Correlators}

\begin{figure}
    \includegraphics[width=0.5\textwidth]{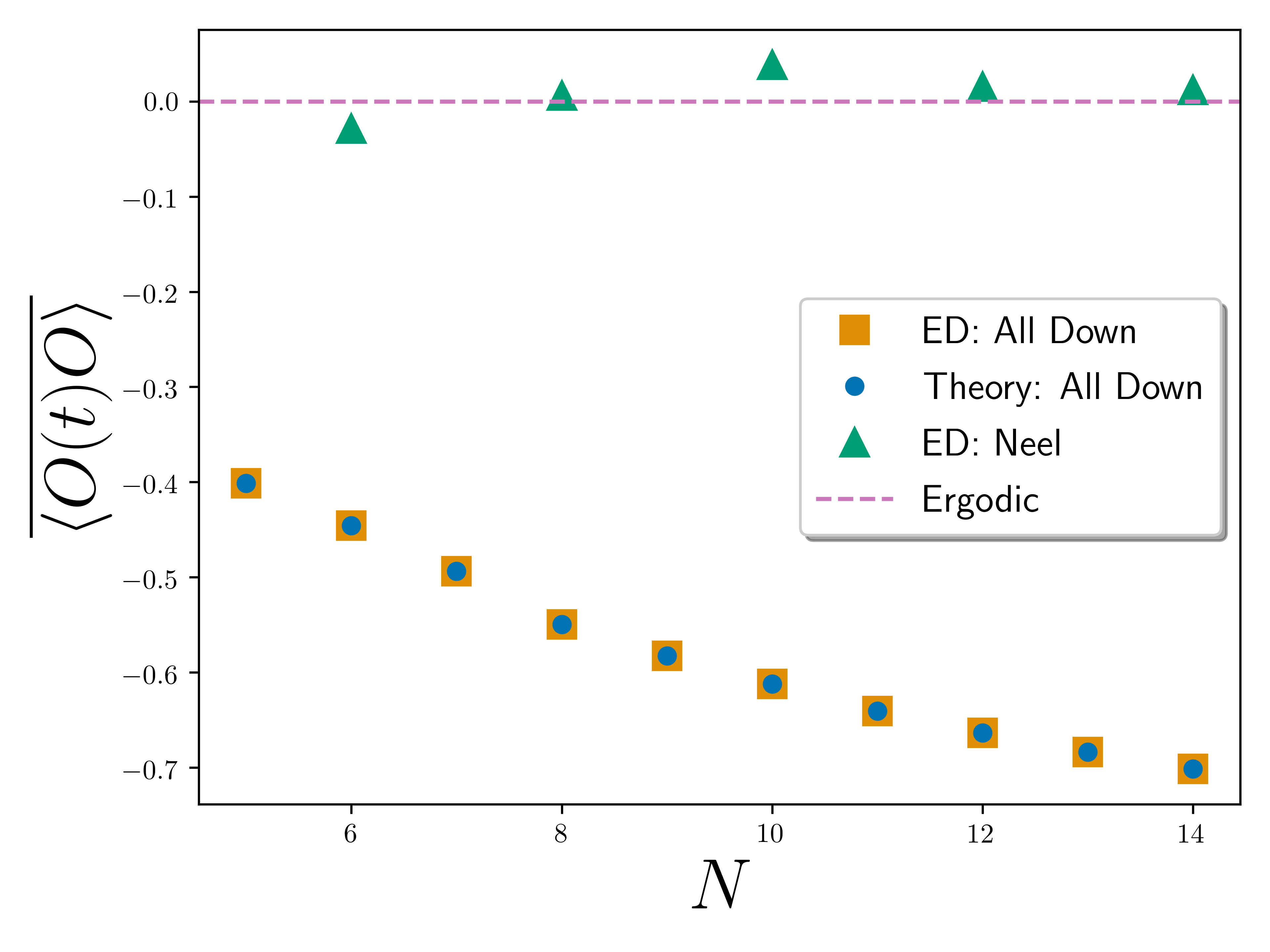}
    \caption{Long time average $\overline{\langle O(t) O \rangle}$ given in Eq. \eqref{eq:non_ergodicity} for the NN-XXX model (Eqs. \eqref{eq:XXX_H_0}, \eqref{eq:XXX_H_I}) for varying number of spins $N$ using exact diagonalization (ED) for initial Neel state (green triangles), and All Down state (yellow squares). Theory values given by Eq. \eqref{eq:F_gsq_full} (blue circles) for all down state, and Neel state matches the ergodic prediction $\overline{\langle O(t) O \rangle}=0$ (dashed line). Even values of Neel state shown such that the initial state has the same number of up and down spins. $B_x = 0.05$, $J = 1$, $J^\prime = 0.8$.}
    \label{fig:XXX_ergodic}
\end{figure}

In this section we will see that the long-time behaviour of the two-point correlators of the correlated quench procedure indeed deviates from the expected factorisation, Eq. \eqref{eq:factorising}. This can be seen using Eq. \eqref{eq:GSQ_ansatz}:
\begin{equation}\label{eq:non_ergodicity}
\begin{split}
\overline{\langle O(t)O\rangle} & = \sum_{\mu\nu}c_\mu(\alpha_0)c_\nu(\alpha_0)O_{\mu\mu}O_{\mu\nu} \\&
= \sum_{\mu\nu}c_\mu(\alpha_0)c_\nu(\alpha_0)(\Delta O c_\mu(\alpha_0)c_\nu(\alpha_0) + O_{\downarrow\downarrow})\\& \qquad \times(\Delta O c_\mu(\alpha_0)c_\nu(\alpha_0) + O_{\downarrow\downarrow}\delta_{\mu\nu})\\&
= \Delta O^2 \textrm{IPR}(|\phi_{\alpha_0}\rangle) + O_{\downarrow\downarrow}^2 \\& \qquad + \Delta O O_{\downarrow\downarrow}(1 + \textrm{IPR}(|\phi_{\alpha_0}\rangle) \\&
\approx O_{\uparrow\uparrow}O_{\downarrow\downarrow} \\&
\neq O_{\downarrow\downarrow}^2,
\end{split}
\end{equation}
where in the penultimate line we have assumed a large effective dimension. This is shown in Fig. \ref{fig:XXX_ergodic} for the spin-chain model described in the next section.

\section{Numerical Model}

For our numerical calculations we use a non-integrable spin-chain model, where the bath is given by the XXX chain with nearest and next-nearest neighbour couplings (NN-XXX). Our system ion is `biased' with a small $B_z$ component. The Hamiltonian is written in the form $H = H_0 + H_I$, with
\begin{equation}\label{eq:XXX_H_0}
\begin{split}
H_0 & = B_z^{(0)}\sigma_z^{(0)} + \sum_{\langle \alpha, \beta\rangle>0} J \boldsymbol{\sigma}_\alpha \cdot \boldsymbol{\sigma}_\beta,
\end{split}
\end{equation}
where we set the system index equal to zero, such that $H_S = B_z^{(0)}\sigma_z^{(0)}$, and,
\begin{equation}\label{eq:XXX_H_I}
H_I = J \boldsymbol{\sigma}_0 \cdot \boldsymbol{\sigma}_{1} + 
\sum_{\langle\langle \alpha, \beta\rangle\rangle} J^\prime \boldsymbol{\sigma}_\alpha \cdot \boldsymbol{\sigma}_\beta,
\end{equation}
where $\boldsymbol{\sigma_\alpha} = ( \sigma_x^{(\alpha)}, \sigma_y^{(\alpha)}, \sigma_z^{(\alpha)})$, and  $\langle \cdots \rangle$ and $\langle\langle \cdots \rangle\rangle$ indicate summations over nearest neighbours and next-nearest neighbours of the respectively. $H_0$ thus describes a system ion placed at one end of the chain, uncoupled from an XXX chain with nearest-neighbour interactions only. The action of $H_I$ is to couple the system ion to both it's neighbour and next-nearest neighbour, as well as include next-nearest neighbour interactions throughout the chain. The system is thus homogeneous, except for a small `bias' field on the system ion only, acting to ensure that the initial state is an excited eigenstate of the conserved quantity $\hat{N}_S = \sigma_z^{(S)}$. The total conserved quantity is the total magnetization $\sum_i \sigma_z^{(i)}$, such that the number of excitations is given by $\hat{N} = \sum_i \frac{1}{2}( \sigma_z^{(i)} + \mathbb{1})$.

This model is chosen for it's resemblance (up to the system $B_z$ field) to that used in Ref. \cite{Iyoda2017}, where the lack of scrambling of quantum information was observed for states with such a conservation law. We argue that Eq. \eqref{eq:GSQ_ansatz} can be seen as the mechanism behind this observation, seeing that scrambling is not violated simply due to a confined subspace by the conservation law, but rather that the mixing of eigenstates in time evolutions is restricted, and thus off-diagonal observable matrix elements may not be treated as random. In Appendix \ref{App:robustness}, we provide similar numerics on a different spin-chain model that breaks this conservation rule.

Our numerical results are shown in Fig. \eqref{fig:XXX}, where we have investigated the NN-XXX model for the observable $\sigma_z^{(S)}$, with initial bath states as the correlated initial state $|\downarrow, \downarrow, \cdots\rangle_B$ (All down), Fig. (\ref{fig:XXX}a), and both the survival probability, $P_0$ and $\sigma_z^{(S)}$ for a highly degenerate product state, $|\uparrow, \downarrow, \cdots\rangle_B$ (Neel), Fig. (\ref{fig:XXX}b), (\ref{fig:XXX}c), respectively. Here we observe that for the correlated initial state the local observable $\sigma_z^{(S)}$ follows the scaling of Eq. \eqref{eq:NE_flucs} as expected from the arguments above. We further see that the survival probability follows this scaling exactly for all initial states.
 
\begin{figure}
    \includegraphics[width=0.5\textwidth]{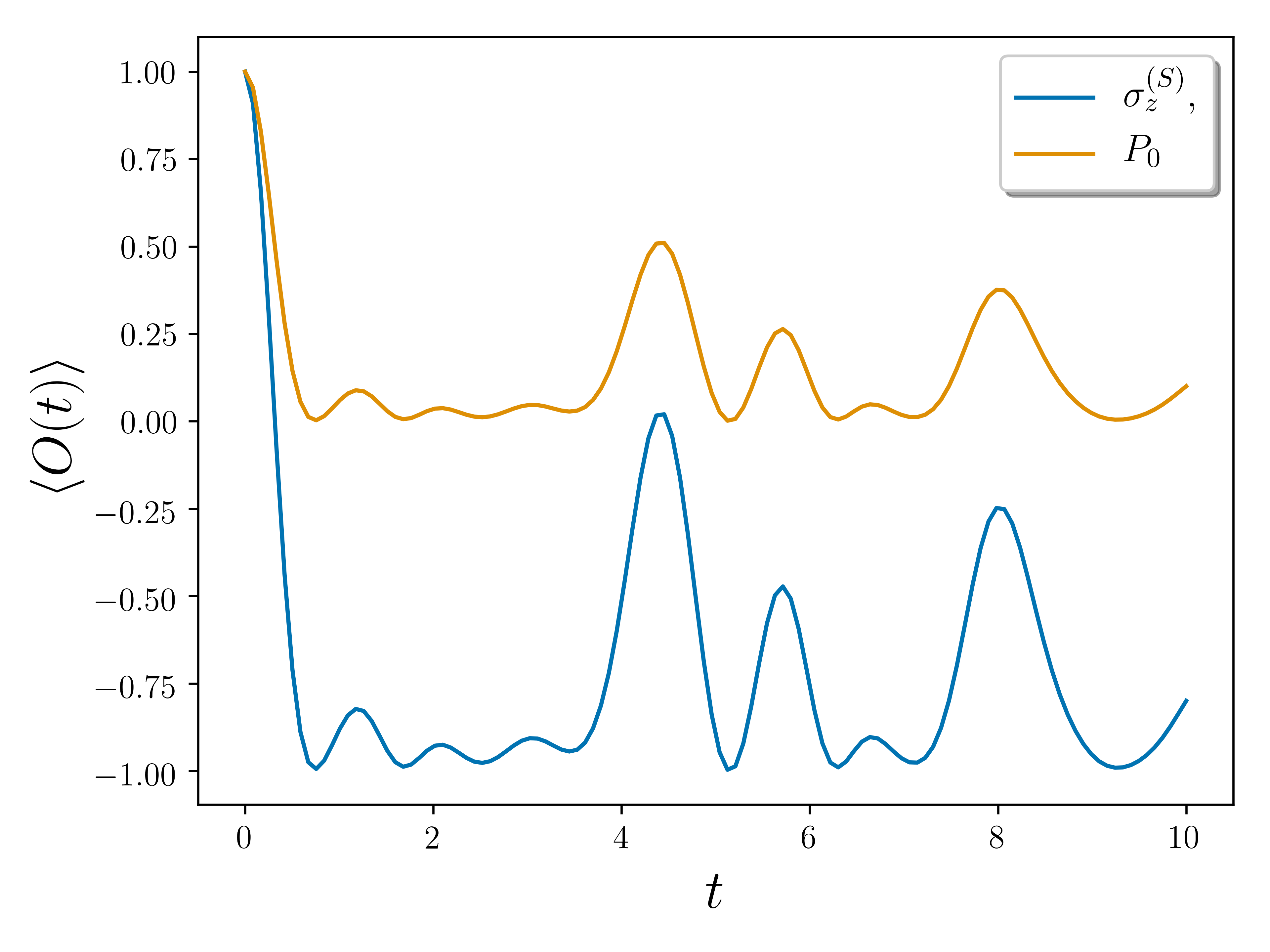}
    \caption{Dynamics of observable $\sigma_z^{(S)}$ (blue) and $P_0$ (orange) for the NN-XXX model (Eqs. \eqref{eq:XXX_H_0}, \eqref{eq:XXX_H_I}), for a correlated quench procedure. We see that the dynamics after a correlated quench closely follows the survival probability, and decays to an equilibrium value $\sim O_{\downarrow\downarrow} = -1$, as expected by Eq. \eqref{eq:time_GS}. $B_z = 0.05, J = 1, J^\prime = 0.8, N = 15$.}
    \label{fig:dynamics}
\end{figure}
\section{Time Evolution} 

The time dependence of an observable $O$ of a closed quantum system initialized in state $|\phi_{\alpha_0}\rangle = |\uparrow\rangle_S|k_0\rangle_B$ may be written as 
\begin{equation}
\begin{split}
\langle O(t) \rangle & = \sum_{\mu\nu} c_\mu(\alpha_0)c_\nu(\alpha_0)e^{-i(E_\mu - E_\nu)t}O_{\mu\nu} \\&
= \Delta O \sum_{\mu\nu}|c_\mu(\alpha_0)|^2|c_\nu(\alpha_0)|^2 e^{-i(E_\mu - E_\nu)t}\\&  + O_{\downarrow\downarrow}\sum_\mu c_\mu^2(k_0, \uparrow),
\end{split}
\end{equation}
where in the second line we have applied Eq. \eqref{eq:GSQ_ansatz}. We thus obtain
\begin{equation}\label{eq:time_GS}
\langle O(t) \rangle = \Delta O P_0(t) + O_{\downarrow\downarrow},
\end{equation}
where $P_0 = |\langle \psi(0)|\psi(t)\rangle|^2 = \sum_{\mu\nu}|c_\mu(\alpha_0)|^2|c_\nu(\alpha_0)|^2e^{-i(E_\mu-E_\nu)t}$ is the survival probability. 

This is corroborated in Fig. \eqref{fig:dynamics}. Note that the case of the correlated quench follows the survival probability dynamics strikingly closely, even faithfully reproducing its fluctuations.

\section{Long-time OTOC}

Exploiting Eq. \eqref{eq:GSQ_ansatz}, the long time OTOC, may be obtained by calculation of Eq. \eqref{eq:F_ave}. 
This is a somewhat tedious calculation, shown in full in Appendix \ref{App:OTOC}. We obtain,
\begin{equation}\label{eq:F_gsq_full}
\begin{split}
\overline{F(t)} & = W_\downarrow^2 V_\downarrow^2 +  W_{\downarrow}^2 \Delta V V_\downarrow + W_\downarrow^2 \Delta V^2 + W_\downarrow^2 \Delta V V_\downarrow \\&
{\cal I}_4[ 4\Delta W W_\downarrow \Delta V V_\downarrow + 2 \Delta W W_\downarrow V_\downarrow^2 \\&
+ 2\Delta W^2 W_\downarrow \Delta V^2 + \Delta W^2 V_\downarrow^2 + \Delta W^2 \Delta V V_\downarrow] \\& 
+ {\cal I}_4^2[2 \Delta W^2 \Delta V V_\downarrow + 2 \Delta W^2 \Delta V^2 ] \\& 
- {\cal I}_8[\Delta W^2 \Delta V^2 + \Delta W^2 \Delta V V_\downarrow ],
\end{split}
\end{equation}
where ${\cal I}_n := \sum_\mu c_\mu^n(\alpha_0)$, and thus ${\cal I}_4 = \textrm{IPR}(|\phi_0\rangle )$. In the limit of large system size, where the IPR is small, this thus simplifies to
\begin{equation}
\begin{split}
\overline{F(t)} \approx W_\downarrow^2 V_\downarrow^2 +  W_{\downarrow}^2 \Delta V V_\downarrow
+ W_\downarrow^2 \Delta V^2 + W_\downarrow^2 \Delta V V_\downarrow ,
\end{split}
\end{equation}
which is equal to unity for $W = V = \sigma_z^{(S)}$, implying that the information on this local observable is \emph{not-scrambled}, even for large non-integrable Hamiltonians. 

\begin{figure}
    \includegraphics[width=0.5\textwidth]{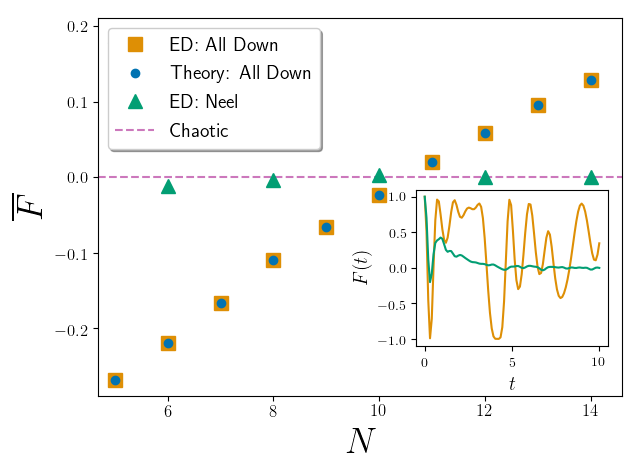}
    \caption{Long time average $\overline{F}$ given in Eq. \eqref{eq:F_ave} for the NN-XXX model (Eqs. \eqref{eq:XXX_H_0}, \eqref{eq:XXX_H_I}) for varying number of spins $N$ using exact diagonalization for initial Neel state (green triangles), and All Down state (yellow squares). Theory values given by Eq. \eqref{eq:F_gsq_full} (blue circles) for all down state, and Neel state matches the chaotic prediction $\overline{F}=0$ (dashed line). Inset shows time evolution of $F(t)$ for $N = 14$ for both Neel (green) and All Down (yellow) states. $B_x = 0.05$, $J = 1$, $J^\prime = 0.8$.}
    \label{fig:XXX_OTOC}
\end{figure}

In Fig. \eqref{fig:XXX_OTOC} we observe the behaviour of the long time value $\overline{F}$ for the initial correlated state. We see that, indeed, $\overline{F}$ does not indicate the scrambling of information, and is exactly in agreement with Eq. \eqref{eq:F_gsq_full}.

This indicates that the dynamics of the system after a correlated quench procedure is non-chaotic. The fact that this system behaves non-chaotically likens the dynamics to that of an integrable system i.e. a system with an extensive number of conserved quantities. Our approach thus provides some intuition on the transition to integrability, upon which an extensive number of initial states will behave as above.

\section{Discussion} 

In this article we have observed and obtained the mechanism behind a scenario in which quantum thermalization occurs in generic chaotic systems, yet other predictions of the ETH are violated due to correlations between the initial state, observable, and a symmetry of the system. The scenario in question we have called a `correlated quench', where the bath is initially in a non-degenerate state with respect to some conservation law. We observe that for given local system observables, thermalization occurs without the full ETH, as the off-diagonal observable elements can be seen to be non-random, as correlations dominate both long and short time behaviour of the observable. Indeed, from the derived expression for off-diagonal observable elements, we have analytically obtained the long-time fluctuations and time evolution of observables, as well as the long time value of the out-of-time ordered correlator. 

The time evolution of observables after a correlated quench follows that of the survival probability closely, which provides a potential method of measuring the survival probability itself. This is useful, for example, as it's Fourier transform is the so-called local density of states (LDOS), or strength function. We see that this may be measured using such a correlated quench protocol.

The arguments outlined above rest on the behaviour of the parameter $\eta$, Eq. \eqref{eq:eta}. This parameter dictates the available states that may mix with the initial state in time evolution. It is thus an important quantity in dictating the ergodicity of the system, or it's ability to scramble quantum information.

C. N. would like to thank C. B. Da\u{g} for enlightening discussions. We acknowledge funding by EPSRC grant no. EP/M508172/1 and project PGC2018-094792-B-I00  (MCIU/AEI/FEDER, UE).

\section{Appendix}
\appendix

\section{Relation to the Rotating Wave Approximation - The Spin-Boson Model} \label{App:RWA}
In this section we give an example of the scaling of time-averaged fluctuations for an integrable spin-boson model. This model serves to provide intuition on the origin of the deviations from the ETH observed in the main text, which will carry over in a straight-forward manner to more general systems. We will see here that the excitation conservation law is equivalent to the Rotating Wave Approximation (RWA) in the familiar model of a single spin coupled to a Bosonic bath.

\begin{figure*}
    \includegraphics[width=0.45\textwidth]{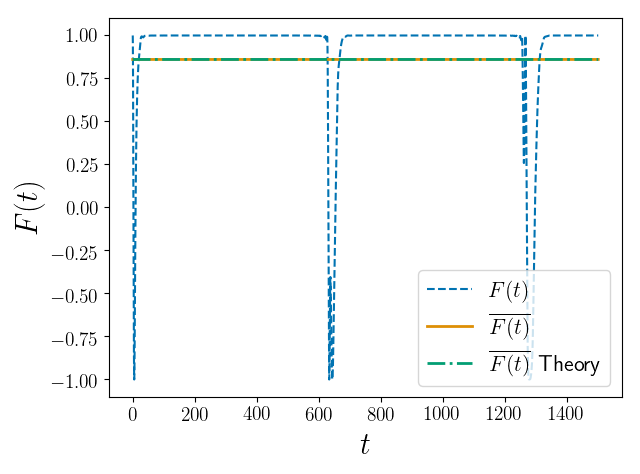} 
    \includegraphics[width=0.45\textwidth]{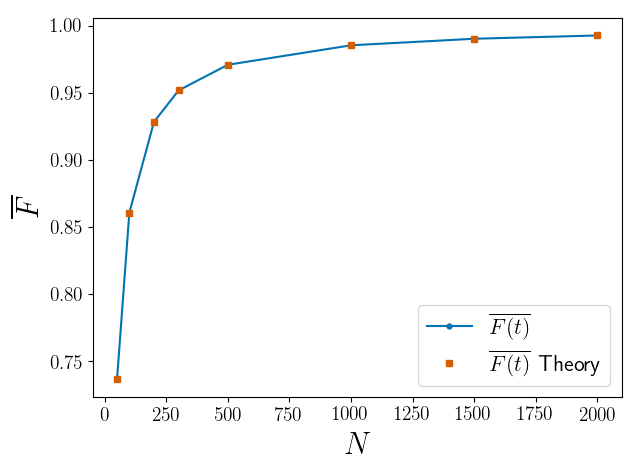} 
    \caption{$F(t)$ for $N = 100$ (left) and $\overline{F(t)}$ (right) for the Spin-Boson Hamiltonian, Eq. \eqref{eq:SB_Hamil}. Theory labels the correlated quench condition result of Eq. \eqref{eq:F_gsq_full}. Note that `revivals' in $F(t)$ significantly contribute to the time average result. $\omega_z = 0.6$, $\omega_n = \frac{n}{N}, \Gamma = \frac{2\pi g^2}{\omega_0} = 0.2$}
    \label{fig:SB_F}
\end{figure*}

The model we discuss is the Spin-Boson Model, which we may write as
\begin{equation}\label{eq:SB_Hamil}
H = H_S + H_B + H_{SB},
\end{equation}
with 
\begin{equation}
H_0 = H_S + H_B = \frac{\omega_z}{2}\sigma_z + \sum_n \omega_n a^\dagger_n a_n,
\end{equation}
and
\begin{equation}
H_{SB} = \sum_n g_n (a_n + a_n^\dagger )(\sigma_+ + \sigma_-),
\end{equation}
where $\sigma_i \ i={x, y, z}$ are the Pauli operators, $\sigma_\pm = \frac{1}{2}(\sigma_x \pm i\sigma_y)$, and $a_n^\dagger(a_n)$ are creation (annihilation) operators of a boson in state $n$.  Making the rotating wave approximation, and thus ignoring counter rotating terms $a_n\sigma_-$ and $a^\dagger\sigma_+$, we obtain
\begin{equation}
H_{SB} = \sum_n g_n (a_n\sigma_+ + a_n^\dagger \sigma_-).
\end{equation}
We can thus see that the total Hamiltonian conserves the total number of excitations $\hat{N} = \frac{1}{2}(\sigma_z + \mathbb{1}) + \sum_n a^\dagger_n a_n$. We thus initiate a correlated quench, such that the initial state is given by $|\uparrow\rangle_S\prod_n|0\rangle_n := |\uparrow, 0\rangle$. As the interaction Hamiltonian with the RWA preserves excitation number, we thus have, that at any later time the state must be some superposition of $|\uparrow, 0\rangle$ and $|\downarrow, 1_n\rangle := |\downarrow\rangle_S|1\rangle_n\prod_{m\neq n} |0\rangle_m$.
\begin{figure}
    \includegraphics[width=0.45\textwidth]{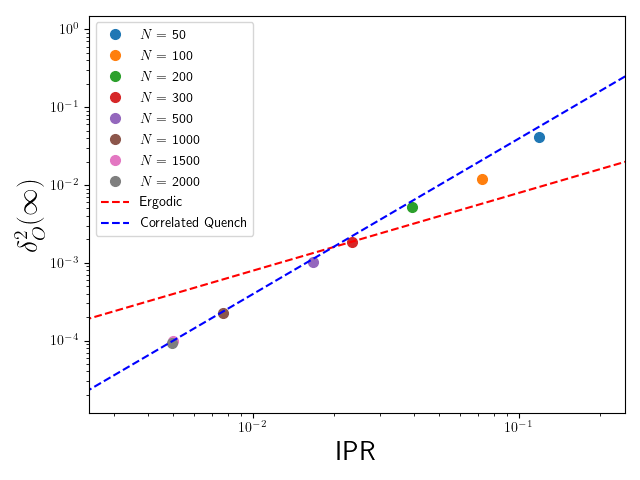}
    \caption{DE Fluctuations versus IPR for Spin-Boson model Eq. \eqref{eq:SB_Hamil} with $\omega_z = 0.6, \Gamma = \frac{2\pi g^2}{\omega_0} = 0.1, \omega_n = n\omega_0 = \frac{n}{N}$. $g_n$ is a random number with mean zero and variance $g$.}
    \label{fig:SB_rand}
\end{figure}
This model is exactly solvable using the Wigner-Weisskopf method \cite{Weisskopf1930, Cohen-Tannoudji} for $g_n = g = $ constant, which gives
\begin{equation}
c_{\uparrow, 0}^{(\mu)} = \frac{g}{(g^2 + \frac{\gamma^2}{4} + E_\mu^2)^{\frac{1}{2}}}, \quad c_{\downarrow, 1_n}^{(\mu)} = \frac{g^2/(E_\mu - \omega_n)}{(g^2 + \frac{\gamma^2}{4} + E_\mu^2)^{\frac{1}{2}}},
\end{equation}
where $\gamma = \frac{2\pi g^2}{\omega_0}$ is the decay rate. Now, the fluctuations and IPR may be easily calculated for this model in the limit $N \to \infty$. Here we can write $\sum_\mu \to \int \frac{dE_\mu}{\omega_0}$, and as in the continuum limit the level spacing $\omega_0 \to 0$, with $\gamma =$ constant, we have $g^2 = \frac{\gamma \omega_0}{2\pi} \to 0 $. We thus have 
\begin{equation}
\textrm{IPR}(|\psi(0)\rangle) = \frac{\gamma^2 \omega_0}{4\pi^2} \int \frac{dE_\mu}{\omega_0} \frac{1}{(\frac{\gamma^2}{4} + E_\mu^2)^2} = \frac{\omega_0}{\pi \gamma}.
\end{equation}
Now, the fluctuations can be found from Eq. \eqref{eq:DE_flucs}. We pick as our observable $\sigma_z$, for which we have $(\sigma_z)_{\mu\nu} = c^{(\mu)}_{\uparrow, 0}c^{(\nu)}_{\uparrow, 0} - \sum_nc^{(\mu)}_{\downarrow, 1_n}c^{(\nu)}_{\downarrow, 1_n} = 2c^{(\mu)}_{\uparrow, 0}c^{(\nu)}_{\uparrow, 0} - \delta_{\mu\nu}$. Thus, Eq \eqref{eq:DE_flucs} may be similarly evaluated to obtain
\begin{equation}\label{eq:flucs_SB}
\begin{split}
\delta_{\sigma_z}^2(\infty) & = 4 \frac{\omega_0^2}{\pi^2\gamma^2} - 10 \frac{\omega_0^3}{\pi^3\gamma^3} \\&
 \approx 4 \textrm{IPR}(|\psi(0)\rangle)^2.
\end{split}
\end{equation}
In Fig. \ref{fig:SB_rand} we plot the DE fluctuations against the IPR for the case where $g_n$ is given by a random number of mean zero and variance $g$, observing that the obtained scaling is still correct for the non-integrable case with random couplings. In Fig. \ref{fig:SB_F} we plot the time dependence and long-time average of $F(t)$, for constant couplings $g_n = g$.

We further note that this case is also exactly fulfilled by the tight-binding model, which may be similarly solved by the Wigner-Weisskopf approach.

\section{Time average of $F(t)$}\label{App:OTOC}
Here we calculate the long-time average of $F(t)$, given by
\begin{equation}\label{eq:F_ave_supp}
\begin{split}
\overline{F} & = \overline{\langle \uparrow, k_0|W^\dagger (t)V^\dagger W(t)V|\uparrow, k_0\rangle}
\\& = \sum_{\mu\nu}c_\mu(k_0, \uparrow)c_\nu(k_0, \uparrow) \overline{\langle \psi_\mu|W^\dagger (t)V^\dagger W(t)V|\psi_\nu\rangle}.
\end{split}
\end{equation}
We first obtain $\overline{F}_{\mu\nu} := \overline{\langle \psi_\mu|W^\dagger (t)V^\dagger W(t)V|\psi_\nu\rangle}$, which can be seen to be equal to
\begin{equation}
\begin{split}
\overline{F}_{\mu_0\nu_0} &= \sum_\mu W_{\mu_0\mu_0}V_{\mu_0\mu}W_{\mu\mu}V_{\mu\nu_0} \\&
+  \sum_\mu W_{\mu_0\mu}V_{\mu\mu}W_{\mu\mu_0}V_{\mu_0\nu_0} - W_{\mu_0\mu_0}^2V_{\mu_0\mu_0}V_{\mu_0\nu_0}\\&
:= \overline{F}_{\mu_0\nu_0}^{(1)} + \overline{F}_{\mu_0\nu_0}^{(2)} - \overline{F}_{\mu_0\nu_0}^{(3)}
\end{split}
\end{equation}

Now, using that $O_{\mu\nu} = \Delta O c_\mu(\uparrow, k_0)c_\nu(\uparrow, k_0) + O_\downarrow \delta_{\mu\nu}$, we may write, using the shorthand $c_\mu(\uparrow, k_0) := c_\mu$,
\begin{equation}
\begin{split}
\overline{F}^{(1)}_{\mu_0\nu_0}& = \sum_\mu (\Delta W c_{\mu_0}c_{\mu_0} + W_\downarrow)(\Delta V c_{\mu_0}c_{\mu} + V_\downarrow \delta_{\mu\mu_0})\\&\times(\Delta W c_{\mu}c_{\mu} + W_\downarrow)(\Delta V c_{\mu}c_{\nu_0} + V_\downarrow \delta_{\mu\nu_0}).
\end{split}
\end{equation}
Performing the expansion in full, we obtain
\begin{equation}
\begin{split}
\overline{F}^{(1)}_{\mu_0\nu_0}& = \sum_\mu \bigg[ W_\downarrow^2 V_\downarrow^2 \delta_{\mu\mu_0} \delta_{\mu\nu_0} + W_\downarrow^2 \Delta V^2 c_\mu^2 c_{\nu_0} c_{\mu_0} \\& + \Delta W^2 V_\downarrow^2 c_\mu^2 c_{\mu_0}^2 \delta_{\mu\mu_0}\delta_{\mu\nu_0} + \Delta W^2 \Delta V^2 c_\mu^4 c_{\nu_0} c_{\mu_0}^3 \\ &
 + \Delta W W_\downarrow \Delta V V_\downarrow ( c_\mu c_{\mu_0}^3 \delta_{\mu\nu_0} + c_\mu^2 c_{\nu_0} \delta_{\mu\mu_0} \\ & \qquad \qquad \qquad + c_\mu^3 c_{\mu_0} \delta_{\mu\nu_0} + c_\mu c_{\nu_0} c_{\mu_0}^2 \delta_{\mu\mu_0} ) \\& 
+\Delta W W_\downarrow \Delta V^2 ( c_\mu^4 c_{\nu_0} c_{\mu_0} + c_\mu^2 c_{\mu_0}^3 c_{\nu_0})\\&
+ \Delta W W_\downarrow V_\downarrow^2 ( c_\mu^2 \delta_{\mu}^2 \delta_{\mu\mu_0} \delta_{\mu\nu_0} + c_{\mu_0}^2 \delta_{\mu\mu_0}\delta_{\mu\nu_0} ) \\& 
+ \Delta W^2 \Delta V V_\downarrow ( c_\mu^3 c_{\mu_0}^2 c_{\nu_0} \delta_{\mu\mu_0} + c_\mu^3 c_{\mu_0}^3) \\& 
+ W_\downarrow^2 \Delta V V_\downarrow ( c_\mu c_{\nu_0} \delta_{\mu\mu_0} + c_\mu c_{\mu_0} \delta_{\mu\nu_0}) \bigg] 
\end{split}
\end{equation}
similarly,
\begin{equation}
\begin{split}
\overline{F}^{(2)}_{\mu_0\nu_0}& = \sum_\mu (\Delta W c_{\mu_0}c_{\mu} + W_\downarrow\delta_{\mu\mu_0})(\Delta V c_{\mu}c_{\mu} + V_\downarrow )\\&\times(\Delta W c_{\mu}c_{\mu_0} + W_\downarrow\delta_{\mu\mu_0})(\Delta V c_{\mu_0}c_{\nu_0} + V_\downarrow \delta_{\mu_0\nu_0}) \\&
= \sum_\mu \bigg[ W_\downarrow^2 V_\downarrow^2 \delta_{\mu_0\nu_0} \delta_{\mu\mu_0} + W_\downarrow^2 \Delta V V_\downarrow c_\mu^2 \delta_{\mu_0\nu_0} \delta_{\mu\mu_0} \\& 
+ 2 \Delta W W_\downarrow V_\downarrow^2 c_\mu c_{\mu_0}  \delta_{\mu_0\nu_0} \delta_{\mu\mu_0} \\ &
+ 2 \Delta W W_\downarrow \Delta V V_\downarrow ( c_\mu^3 c_{\mu_0}  \delta_{\mu_0\nu_0} \delta_{\mu\mu_0} + c_\mu c_{\mu_0}^2 c_{\nu_0} \delta_{\mu\mu_0} ) \\&
+ W_\downarrow^2 \Delta V V_\downarrow c_{\mu_0} c_{\nu_0} \delta_{\mu\mu_0} + W_\downarrow^2 \Delta V^2 c_{\mu_0} c_{\nu_0} \delta_{\mu\mu_0} \\&
+ \Delta W^2 V_\downarrow^2 c_\mu^2 c_{\mu_0}^2 \delta_{\mu_0 \nu_0} + \Delta W^2 \Delta V^2 c_\mu^4 c_{\mu_0}^3 c_{\nu_0} \\& 
+ 2 \Delta W W_\downarrow \Delta V^2 c_\mu^3 c_{\mu_0}^2 c_{\nu_0} \delta_{\mu\mu_0}  \\ &
+ \Delta W^2 \Delta V V_\downarrow ( c_\mu^2 c_{\mu_0}^3 c_{\nu_0} + c_\mu^4 c_{\mu_0}^2 \delta_{\mu_0 \nu_0}) \bigg]
\end{split}
\end{equation}
and,
\begin{equation}
\begin{split}
\overline{F}^{(3)}_{\mu_0\nu_0} & = (\Delta W c_{\mu_0}c_{\mu_0} + W_\downarrow)(\Delta V c_{\mu_0}c_{\mu_0} + V_\downarrow )\\&\times(\Delta W c_{\mu_0}c_{\mu_0} + W_\downarrow)(\Delta V c_{\mu_0}c_{\nu_0} + V_\downarrow \delta_{\mu_0\nu_0}) \\&
= W_\downarrow^2 V_\downarrow^2 \delta_{\mu_0\nu_0} + W_\downarrow^2 \Delta V V_\downarrow c_{\mu_0} \delta_{\mu_0 \nu_0} \\& 
+ 2 \Delta W W_\downarrow V_\downarrow^2 c_{\mu_0} \delta_{\mu_0 \nu_0} + W_\downarrow^2 \Delta V^2 c_{\mu_0}^3 c_{\nu_0} \\& 
+ 2 \Delta W W_\downarrow \Delta V V_\downarrow (c_{\mu_0}^3 c_{\nu_0} + c_{\mu_0}^4\delta_{\mu_0 \nu_0} ) \\&
+ \Delta W^2 V_\downarrow^2 c_{\mu_0}^4 \delta_{\mu_0 \nu_0} + 2 \Delta W W_\downarrow \Delta V^2 c_{\mu_0}^5 c_{\nu_0} \\& + \Delta W^2 \Delta V V_\downarrow c_{\mu_0}^5 c_{\nu_0} + \Delta W^2 \Delta V V_\downarrow c_{\mu_0}^6 \delta_{\mu_0 \nu_0} \\&
 + \Delta W^2 \Delta V^2 c_{\mu_0}^7 c_{\nu_0} \bigg]
\end{split}
\end{equation}
Now, using that,
\begin{equation}
\overline{F(t)} = \sum_{\mu_0 \nu_0} c_{\mu_0} c_{\nu_0} \overline{F}_{\mu_0 \nu_0}, 
\end{equation}
and defining
\begin{equation}
{\cal I}_n = \sum_\mu c_\mu^n,
\end{equation}
we thus obtain (noting that ${\cal I}_2 = 1$),
\begin{equation}\label{eq:F_gsq_full_2}
\begin{split}
\overline{F(t)} & = W_\downarrow^2 V_\downarrow^2 +  W_{\downarrow}^2 \Delta V V_\downarrow + W_\downarrow^2 \Delta V^2 + W_\downarrow^2 \Delta V V_\downarrow \\&
+ {\cal I}_4[ 4\Delta W W_\downarrow \Delta V V_\downarrow + 2 \Delta W W_\downarrow V_\downarrow^2 \\&
+ 2\Delta W^2 W_\downarrow \Delta V^2 + \Delta W^2 V_\downarrow^2 + \Delta W^2 \Delta V V_\downarrow] \\& 
+ {\cal I}_4^2[2 \Delta W^2 \Delta V V_\downarrow + 2 \Delta W^2 \Delta V^2 ] \\& 
- {\cal I}_8[\Delta W^2 \Delta V^2 + \Delta W^2 \Delta V V_\downarrow ],
\end{split}
\end{equation}
which is the result shown in the main text. We note that terms in ${\cal I}_n$ can be seen as finite size effects, which become negligible as $N \to \infty$. Indeed ${\cal I}_4$ is equal to the inverse participation ratio. For large system sizes, we thus expect,
\begin{equation}
\begin{split}
\overline{F(t)} \approx W_\downarrow^2 V_\downarrow^2 +  W_{\downarrow}^2 \Delta V V_\downarrow
+ W_\downarrow^2 \Delta V^2 + W_\downarrow^2 \Delta V V_\downarrow
\end{split}
\end{equation}
We use the full Equation in the numerics, however, as the finite size effects can be seen to be important in the system sizes studied numerically.

\section{Discussion of larger system sizes}\label{App:Larger_System}
In the main text we focused on the case where $H_S$ is a $2\times 2$ Hamiltonian matrix. Here we show that in certain conditions the main arguments similarly follow for larger systems, of Hilbert space dimension ${\cal N}_S$, with eigenstates $\{ |s\rangle_S \}_{s = 1, ..., {\cal N}_S}$. We now write the initial state as $|\psi(0)\rangle = |s_i\rangle_S |k_0\rangle_B$, where $|k_0\rangle_B$ is again a non-degenerate state of some conserved quantity, and $|s_i\rangle_S$ is the initial (excited) system state.
\begin{figure*}
    \includegraphics[width=\textwidth]{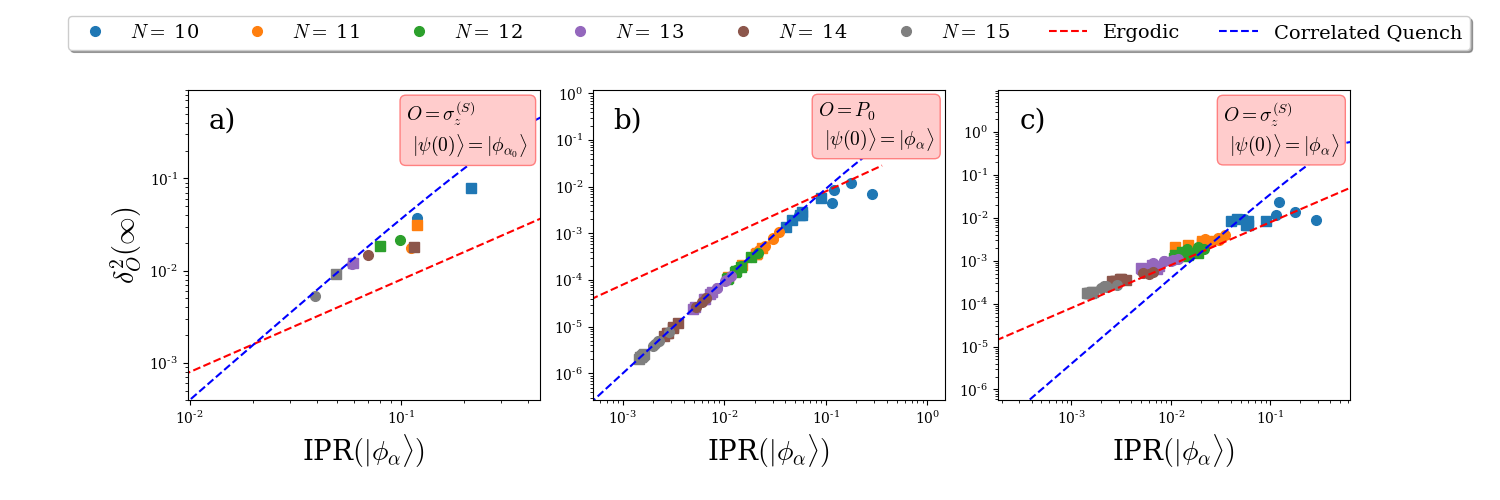}
    \caption{DE Fluctuations versus IPR for Hamiltonian Eqs. \eqref{eq:H_S_Chain}, \eqref{eq:H_B}, and \eqref{eq:H_SB_Chain}. $B_x^{(S)} = 0, B_x^{(B)} = 0.3, B_z^{(S)} = 0.8, B_z^{(B)} = 0, J_x^{(B)} = 1, J_z^{(B)} = 0.1$. Coupling strengths: $J^{(SB)}_x = 0.8, 1.0$ for circles and squares respectively. Initial state is $|\uparrow\rangle_S |\psi_{\alpha_0}\rangle_B$, where $|\psi_{\alpha_0}\rangle_B$ is the ground state of the bath Hamiltonian $H_B$. $N_m = 5$, such that the chain is non-integrable.}
    \label{fig:SC_flucs}
\end{figure*}

We can show this simply by deriving Eq. \eqref{eq:GSQ_ansatz} for an arbitrary size $H_S$. Indeed, we can do this by following the same approach as the main text, with some additional requirements. Writing instead
\begin{equation}
|\psi_\mu \rangle = \sum_k c_\mu(k, s_i)|s_i\rangle_S |k\rangle_B + \sum_{s\neq s_i}^{{\cal N}_S} \sum_k c_\mu(k, s)|s\rangle_S |k\rangle_B.
\end{equation}
Now, once again, if the system observable is diagonal in the non-interacting basis, we have $O_{s s^\prime} := \langle s|O|s^\prime \rangle \sim \delta_{ss^\prime}$, such that
\begin{equation}
\begin{split}
O_{\mu\nu} & = \sum_k c_\mu(s_i, k)c_\nu(s_i, k)O_{s_i s_i}  \\& \qquad\qquad\qquad + \sum_{s\neq s_i}\sum_k c_\mu(s, k)c_\nu(s, k)O_{ss} \\ &
= \sum_k c_\mu(s_i, k)c_\nu(s_i, k)O_{s_i s_i} \\& \qquad\qquad + \sum_{s\neq s_i}O_{ss} \langle\psi_\mu| \sum_k |s\rangle_S|k\rangle_B{}_B\langle k |{}_S\langle s|\psi_\nu \rangle \\&
= \sum_k c_\mu(s_i, k)c_\nu(s_i, k)O_{s_i s_i} \\& \qquad+ \sum_{s\neq s_i}O_{ss} \langle\psi_\mu|( \mathbb{1} -\sum_{s^\prime \neq s} \sum_k |s^\prime\rangle_S|k\rangle_B{}_B\langle k |{}_S\langle s^\prime|)|\psi_\nu \rangle \\&
= c_\mu(s_i, k_0)c_\nu(s_i, k_0)O_{s_i s_i} - O_{s_i s_i}\delta_{\mu\nu} \\& \qquad\qquad - \sum_{s\neq s_i}O_{ss} \sum_{s^\prime \neq s} \sum_k c_\mu(s^\prime, k)c_\nu(s^\prime, k),
\end{split}
\end{equation}
where we have used that $O$ may be taken as traceless, such that $\sum_{s\neq s_i}O_{ss} = - O_{s_i s_i}$.
Now, we see that the same form is recovered up to a correction given by the last term. We see that this term is dictated by the quantity $\sum_k c_\mu(s, k)c_\nu(s, k)$. Assuming conservation of excitation number, and thus this summation is given by 
\begin{equation}
 \sum_{s^\prime \neq s} \sum_k c_\mu(s^\prime, k)c_\nu(s^\prime, k) = \sum_{\Delta s \neq 0} c_\mu(s_0 - \Delta s, k_0 + \Delta s),
\end{equation}
where $\Delta s = s_i - s^\prime$ is the change in excitation number of the system. Note that $\Delta s$ is always positive when $|k_0\rangle_B$ is initialized with zero excitations, or if the $s_i = \max s$. We then have the correction term as equal to
\begin{equation}
-\sum_{s \neq s_i}O_{ss}\sum_{s^\prime \neq s} c_\mu(s^\prime, k_0 + s_i - s^\prime)c_\nu(s^\prime, k_0 + s_i - s^\prime).
\end{equation}
For high ${\cal N}_S$, then, this term can dominate the off-diagonal contribution. We note, however, that for transnationally invariant eigenstates, as expected for a many-body quantum system away from the edges (which we note the total system + bath state fulfils, given that the initial system energy is large enough), we can write $c_\mu(s, k) \approx c_\mu(s + k)$, and thus the correction term becomes,
\begin{equation}
\begin{split}
O_{s_i s_i}({\cal N}_S - 1) c_\mu(k_0, s_i)c_\nu(k_0, s_i),
\end{split}
\end{equation}
where we have once again used that $O$ may be taken as traceless. We then have,
\begin{equation}
O_{\mu\nu} \approx O_{s_i s_i} {\cal N}_S c_\mu(k_0 + s)c_\nu(k_0 + s) - O_{s_i s_i}\delta_{\mu\nu},
\end{equation}
which is of the form of Eq. \eqref{eq:GSQ_ansatz}.

\section{Robustness to multiple excitations}\label{App:robustness}

\begin{figure}
    \includegraphics[width=0.5\textwidth]{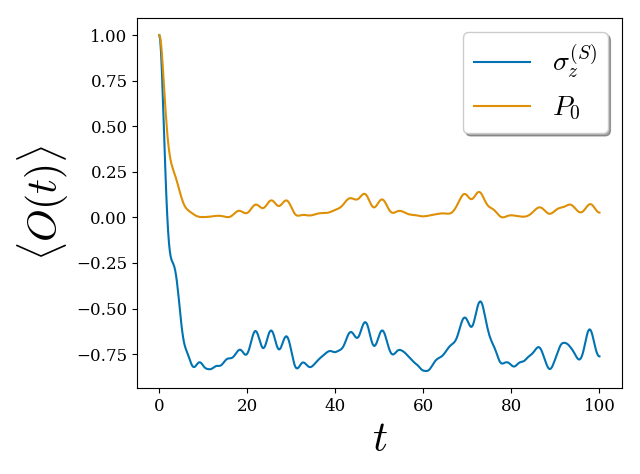}
    \caption{Dynamics of observable $\sigma_z^{(S)}$ (blue) and $P_0$ (orange) for Hamiltonian Eqs. \eqref{eq:H_S_Chain}, \eqref{eq:H_B}, and \eqref{eq:H_SB_Chain}. $B_x^{(S)} = 0, B_x^{(B)} = 0.3, B_z^{(S)} = 0.8, B_z^{(B)} = 0, J_x^{(B)} = 1, J_z^{(B)} = 0.1$. Coupling strengths: $J^{(SB)}_x = 0.8, 1.0$ for circles and squares respectively. Initial state is $|\uparrow\rangle_S |\psi_{\alpha_0}\rangle_B$, where $|\psi_{\alpha_0}\rangle_B$ is the ground state of the bath Hamiltonian $H_B$. $N_m = 5, N = 14$.}
    \label{fig:SC_times}
\end{figure}

In this section we show that in some cases Eq. \eqref{eq:GSQ_ansatz} may be applied outside the regime where it is exact, that is, to the case where there is no conservation law that we may use as our excitation number. We will see that when the initial bath state is the ground state of $H_B$, Eq. \eqref{eq:GSQ_ansatz} is approximately fulfilled due to the energy cost associated to an excited state of the bath.

We will observe this using a different Spin-Chain Hamiltonian, of the form $H = H_S + H_B + H_{SB}$, with
\begin{equation}\label{eq:H_S_Chain}
H_S = B_{z}^{(S)}\sigma_z^{(1)}
\end{equation}
where $\{\sigma_i^{(n)}\}\quad i = {x, y, z}$ are the Pauli operators acting on site $n$. The bath Hamiltonian is a spin-chain of length $N$, with nearest-neighbour Ising and XX interactions subjected to both $B_z$ and $B_x$ fields
\begin{equation}\label{eq:H_B}
\begin{split}
& H_B =  \sum_{n > 1}^N( B_{z}^{(B)}\sigma_z^{(n)}  + B_{x}^{(B)}\sigma_x^{(n)} +\\ & \sum_{n > 1}^{N-1}J_z\sigma_z^{(n)}\sigma_z^{(n+1)} + J_x(\sigma_+^{(n)}\sigma_-^{(n+1)} + \sigma_-^{(n)}\sigma_+^{(n+1)} )).
\end{split}
\end{equation}
The interaction part of the Hamiltonian is given by,
\begin{equation}\label{eq:H_SB_Chain}
H_{SB} = 
J_x^{(SB)}(\sigma_+^{(1)}\sigma_-^{N_{\rm m}} + \sigma_-^{(1)}\sigma_+^{N_{\rm m}}),
\end{equation}
where we use $N_{\rm m} = 5$ throughout, such that the bath is described by a 1-D chain with indices $2, \cdots, N$, and the system (site 1) is coupled to a single spin at site 5. 

In this case, we have that there is no conservation of excitation number, so Eq. \eqref{eq:GSQ_ansatz} is at best approximate. In Fig. \ref{fig:SC_flucs} we plot the scaling of fluctuations for this Hamiltonian for two initial states. In the first, we have the system initialized in the excited state $|\uparrow\rangle_S$, and the bath initialized in the ground state of $H_B$, and for the second, we choose the same system state, and a random mid-energy eigenstate of the bath. We observe that  for the bath ground state initial state, Fig. \ref{fig:SC_flucs}a), the correlated quench results are still a good approximation, however this gets worse as the systems size increases, due to the presence of more states that are close to the ground state with multiple excitations, that are able to be excited. As above, we see that the fluctuations in survival probability differ from the ETH prediction, Fig. \ref{fig:SC_flucs}b), for all initial states, and the random mid-energy eigenstates scale according to the ETH, Fig. \ref{fig:SC_flucs}c).

We thus observe that the implied scaling is robust in some cases to the presence of non-excitation number conserving terms. This can be attributed to a similar mechanism, where due to conservation of energy, if the system is measured to be in the excited state, the bath is (at least likely to be) in the ground state, and thus the system observable is equivalent to the survival probability. This can be seen in the case of time evolution, see Fig. \ref{fig:SC_times}, where we see that the description of diagonal local systems observables in terms of the survival probability remains a good approximation.

\bibliographystyle{apsrev4-1}
\input{Arxiv_Submission.bbl}

\end{document}

%% file: Arxiv_Submission.bbl
%

%% file: Arxiv_Submission.bbl
\begin{thebibliography}{64}%
\makeatletter
\providecommand \@ifxundefined [1]{%
 \@ifx{#1\undefined}
}%
\providecommand \@ifnum [1]{%
 \ifnum #1\expandafter \@firstoftwo
 \else \expandafter \@secondoftwo
 \fi
}%
\providecommand \@ifx [1]{%
 \ifx #1\expandafter \@firstoftwo
 \else \expandafter \@secondoftwo
 \fi
}%
\providecommand \natexlab [1]{#1}%
\providecommand \enquote  [1]{``#1''}%
\providecommand \bibnamefont  [1]{#1}%
\providecommand \bibfnamefont [1]{#1}%
\providecommand \citenamefont [1]{#1}%
\providecommand \href@noop [0]{\@secondoftwo}%
\providecommand \href [0]{\begingroup \@sanitize@url \@href}%
\providecommand \@href[1]{\@@startlink{#1}\@@href}%
\providecommand \@@href[1]{\endgroup#1\@@endlink}%
\providecommand \@sanitize@url [0]{\catcode `\\12\catcode `\$12\catcode
  `\&12\catcode `\#12\catcode `\^12\catcode `\_12\catcode `\%12\relax}%
\providecommand \@@startlink[1]{}%
\providecommand \@@endlink[0]{}%
\providecommand \url  [0]{\begingroup\@sanitize@url \@url }%
\providecommand \@url [1]{\endgroup\@href {#1}{\urlprefix }}%
\providecommand \urlprefix  [0]{URL }%
\providecommand \Eprint [0]{\href }%
\providecommand \doibase [0]{http://dx.doi.org/}%
\providecommand \selectlanguage [0]{\@gobble}%
\providecommand \bibinfo  [0]{\@secondoftwo}%
\providecommand \bibfield  [0]{\@secondoftwo}%
\providecommand \translation [1]{[#1]}%
\providecommand \BibitemOpen [0]{}%
\providecommand \bibitemStop [0]{}%
\providecommand \bibitemNoStop [0]{.\EOS\space}%
\providecommand \EOS [0]{\spacefactor3000\relax}%
\providecommand \BibitemShut  [1]{\csname bibitem#1\endcsname}%
\let\auto@bib@innerbib\@empty
\bibitem [{\citenamefont {{Rigol}}\ \emph {et~al.}(2008)\citenamefont
  {{Rigol}}, \citenamefont {{Dunjko}},\ and\ \citenamefont
  {{Olshanii}}}]{Rigol2008}%
  \BibitemOpen
  \bibfield  {author} {\bibinfo {author} {\bibfnamefont {M.}~\bibnamefont
  {{Rigol}}}, \bibinfo {author} {\bibfnamefont {V.}~\bibnamefont {{Dunjko}}}, \
  and\ \bibinfo {author} {\bibfnamefont {M.}~\bibnamefont {{Olshanii}}},\
  }\href {\doibase 10.1038/nature06838} {\bibfield  {journal} {\bibinfo
  {journal} {\nat}\ }\textbf {\bibinfo {volume} {452}},\ \bibinfo {pages} {854}
  (\bibinfo {year} {2008})}\BibitemShut {NoStop}%
\bibitem [{\citenamefont {Yukalov}(2011)}]{Yukalov2011}%
  \BibitemOpen
  \bibfield  {author} {\bibinfo {author} {\bibfnamefont {V.}~\bibnamefont
  {Yukalov}},\ }\href {\doibase 10.1002/lapl.201110002} {\bibfield  {journal}
  {\bibinfo  {journal} {Laser Physics Letters}\ }\textbf {\bibinfo {volume}
  {8}},\ \bibinfo {pages} {485} (\bibinfo {year} {2011})}\BibitemShut {NoStop}%
\bibitem [{\citenamefont {Eisert}\ \emph {et~al.}(2014)\citenamefont {Eisert},
  \citenamefont {Friesdorf},\ and\ \citenamefont {Gogolin}}]{Eisert2015}%
  \BibitemOpen
  \bibfield  {author} {\bibinfo {author} {\bibfnamefont {J.}~\bibnamefont
  {Eisert}}, \bibinfo {author} {\bibfnamefont {M.}~\bibnamefont {Friesdorf}}, \
  and\ \bibinfo {author} {\bibfnamefont {C.}~\bibnamefont {Gogolin}},\ }\href
  {http://arxiv.org/abs/1408.5148{\%}0Ahttp://dx.doi.org/10.1038/nphys3215}
  {\bibfield  {journal} {\bibinfo  {journal} {Nature Physics}\ }\textbf
  {\bibinfo {volume} {11}},\ \bibinfo {pages} {124} (\bibinfo {year}
  {2014})}\BibitemShut {NoStop}%
\bibitem [{\citenamefont {D'Alessio}\ \emph {et~al.}(2016)\citenamefont
  {D'Alessio}, \citenamefont {Kafri}, \citenamefont {Polkovnikov},\ and\
  \citenamefont {Rigol}}]{DAlessio2016}%
  \BibitemOpen
  \bibfield  {author} {\bibinfo {author} {\bibfnamefont {L.}~\bibnamefont
  {D'Alessio}}, \bibinfo {author} {\bibfnamefont {Y.}~\bibnamefont {Kafri}},
  \bibinfo {author} {\bibfnamefont {A.}~\bibnamefont {Polkovnikov}}, \ and\
  \bibinfo {author} {\bibfnamefont {M.}~\bibnamefont {Rigol}},\ }\href
  {\doibase 10.1080/00018732.2016.1198134} {\bibfield  {journal} {\bibinfo
  {journal} {Advances in Physics}\ }\textbf {\bibinfo {volume} {65}},\ \bibinfo
  {pages} {239} (\bibinfo {year} {2016})}\BibitemShut {NoStop}%
\bibitem [{\citenamefont {Gogolin}\ and\ \citenamefont
  {Eisert}(2016)}]{Gogolin2016}%
  \BibitemOpen
  \bibfield  {author} {\bibinfo {author} {\bibfnamefont {C.}~\bibnamefont
  {Gogolin}}\ and\ \bibinfo {author} {\bibfnamefont {J.}~\bibnamefont
  {Eisert}},\ }\href {\doibase 10.1088/0034-4885/79/5/056001} {\bibfield
  {journal} {\bibinfo  {journal} {Reports on Progress in Physics}\ }\textbf
  {\bibinfo {volume} {79}},\ \bibinfo {pages} {056001} (\bibinfo {year}
  {2016})}\BibitemShut {NoStop}%
\bibitem [{\citenamefont {Borgonovi}\ \emph {et~al.}(2016)\citenamefont
  {Borgonovi}, \citenamefont {Izrailev}, \citenamefont {Santos},\ and\
  \citenamefont {Zelevinsky}}]{Borgonovi2016}%
  \BibitemOpen
  \bibfield  {author} {\bibinfo {author} {\bibfnamefont {F.}~\bibnamefont
  {Borgonovi}}, \bibinfo {author} {\bibfnamefont {F.~M.}\ \bibnamefont
  {Izrailev}}, \bibinfo {author} {\bibfnamefont {L.~F.}\ \bibnamefont
  {Santos}}, \ and\ \bibinfo {author} {\bibfnamefont {V.~G.}\ \bibnamefont
  {Zelevinsky}},\ }\bibfield  {booktitle} {\emph {\bibinfo {booktitle} {Physics
  Reports}},\ }\href {\doibase 10.1016/j.physrep.2016.02.005} {\ \textbf
  {\bibinfo {volume} {626}},\ \bibinfo {pages} {1} (\bibinfo {year} {2016})},\
  \Eprint {http://arxiv.org/abs/1602.01874} {arXiv:1602.01874} \BibitemShut
  {NoStop}%
\bibitem [{\citenamefont {Gring}\ \emph {et~al.}(2012)\citenamefont {Gring},
  \citenamefont {Kuhnert}, \citenamefont {Langen}, \citenamefont {Kitagawa},
  \citenamefont {Rauer}, \citenamefont {Schreitl}, \citenamefont {Mazets},
  \citenamefont {{Adu Smith}}, \citenamefont {Demler},\ and\ \citenamefont
  {Schmiedmayer}}]{Gring2012}%
  \BibitemOpen
  \bibfield  {author} {\bibinfo {author} {\bibfnamefont {M.}~\bibnamefont
  {Gring}}, \bibinfo {author} {\bibfnamefont {M.}~\bibnamefont {Kuhnert}},
  \bibinfo {author} {\bibfnamefont {T.}~\bibnamefont {Langen}}, \bibinfo
  {author} {\bibfnamefont {T.}~\bibnamefont {Kitagawa}}, \bibinfo {author}
  {\bibfnamefont {B.}~\bibnamefont {Rauer}}, \bibinfo {author} {\bibfnamefont
  {M.}~\bibnamefont {Schreitl}}, \bibinfo {author} {\bibfnamefont
  {I.}~\bibnamefont {Mazets}}, \bibinfo {author} {\bibfnamefont
  {D.}~\bibnamefont {{Adu Smith}}}, \bibinfo {author} {\bibfnamefont
  {E.}~\bibnamefont {Demler}}, \ and\ \bibinfo {author} {\bibfnamefont
  {J.}~\bibnamefont {Schmiedmayer}},\ }\href {\doibase 10.1126/science.1224953}
  {\bibfield  {journal} {\bibinfo  {journal} {Science}\ }\textbf {\bibinfo
  {volume} {337}},\ \bibinfo {pages} {1318} (\bibinfo {year}
  {2012})}\BibitemShut {NoStop}%
\bibitem [{\citenamefont {Schneider}\ \emph {et~al.}(2012)\citenamefont
  {Schneider}, \citenamefont {Porras},\ and\ \citenamefont
  {Schaetz}}]{Schneider2012}%
  \BibitemOpen
  \bibfield  {author} {\bibinfo {author} {\bibfnamefont {C.}~\bibnamefont
  {Schneider}}, \bibinfo {author} {\bibfnamefont {D.}~\bibnamefont {Porras}}, \
  and\ \bibinfo {author} {\bibfnamefont {T.}~\bibnamefont {Schaetz}},\ }\href
  {\doibase 10.1088/0034-4885/75/2/024401} {\bibfield  {journal} {\bibinfo
  {journal} {Rep. Prog. Phys.}\ }\textbf {\bibinfo {volume} {75}},\ \bibinfo
  {pages} {24401} (\bibinfo {year} {2012})}\BibitemShut {NoStop}%
\bibitem [{\citenamefont {Georgescu}\ \emph {et~al.}(2014)\citenamefont
  {Georgescu}, \citenamefont {Ashhab},\ and\ \citenamefont
  {Nori}}]{Georgescu2014}%
  \BibitemOpen
  \bibfield  {author} {\bibinfo {author} {\bibfnamefont {I.~M.}\ \bibnamefont
  {Georgescu}}, \bibinfo {author} {\bibfnamefont {S.}~\bibnamefont {Ashhab}}, \
  and\ \bibinfo {author} {\bibfnamefont {F.}~\bibnamefont {Nori}},\ }\href
  {\doibase 10.1103/RevModPhys.86.153} {\bibfield  {journal} {\bibinfo
  {journal} {Rev. Mod. Phys.}\ }\textbf {\bibinfo {volume} {86}},\ \bibinfo
  {pages} {153} (\bibinfo {year} {2014})}\BibitemShut {NoStop}%
\bibitem [{\citenamefont {Schreiber}\ \emph {et~al.}(2015)\citenamefont
  {Schreiber}, \citenamefont {Hodgman}, \citenamefont {Bordia}, \citenamefont
  {L{\"{u}}schen}, \citenamefont {Fischer}, \citenamefont {Vosk}, \citenamefont
  {Altman}, \citenamefont {Schneider},\ and\ \citenamefont
  {Bloch}}]{Schreiber2015}%
  \BibitemOpen
  \bibfield  {author} {\bibinfo {author} {\bibfnamefont {M.}~\bibnamefont
  {Schreiber}}, \bibinfo {author} {\bibfnamefont {S.~S.}\ \bibnamefont
  {Hodgman}}, \bibinfo {author} {\bibfnamefont {P.}~\bibnamefont {Bordia}},
  \bibinfo {author} {\bibfnamefont {H.~P.}\ \bibnamefont {L{\"{u}}schen}},
  \bibinfo {author} {\bibfnamefont {M.~H.}\ \bibnamefont {Fischer}}, \bibinfo
  {author} {\bibfnamefont {R.}~\bibnamefont {Vosk}}, \bibinfo {author}
  {\bibfnamefont {E.}~\bibnamefont {Altman}}, \bibinfo {author} {\bibfnamefont
  {U.}~\bibnamefont {Schneider}}, \ and\ \bibinfo {author} {\bibfnamefont
  {I.}~\bibnamefont {Bloch}},\ }\href {\doibase 10.1126/science.aaa7432}
  {\bibfield  {journal} {\bibinfo  {journal} {Science}\ }\textbf {\bibinfo
  {volume} {349}},\ \bibinfo {pages} {842} (\bibinfo {year}
  {2015})}\BibitemShut {NoStop}%
\bibitem [{\citenamefont {Neill}\ \emph {et~al.}(2016)\citenamefont {Neill},
  \citenamefont {Roushan}, \citenamefont {Fang}, \citenamefont {Chen},
  \citenamefont {Kolodrubetz}, \citenamefont {Chen}, \citenamefont {Megrant},
  \citenamefont {Barends}, \citenamefont {Campbell}, \citenamefont {Chiaro},
  \citenamefont {Dunsworth}, \citenamefont {Jeffrey}, \citenamefont {Kelly},
  \citenamefont {Mutus}, \citenamefont {O'Malley}, \citenamefont {Quintana},
  \citenamefont {Sank}, \citenamefont {Vainsencher}, \citenamefont {Wenner},
  \citenamefont {White}, \citenamefont {Polkovnikov},\ and\ \citenamefont
  {Martinis}}]{Neill2016}%
  \BibitemOpen
  \bibfield  {author} {\bibinfo {author} {\bibfnamefont {C.}~\bibnamefont
  {Neill}}, \bibinfo {author} {\bibfnamefont {P.}~\bibnamefont {Roushan}},
  \bibinfo {author} {\bibfnamefont {M.}~\bibnamefont {Fang}}, \bibinfo {author}
  {\bibfnamefont {Y.}~\bibnamefont {Chen}}, \bibinfo {author} {\bibfnamefont
  {M.}~\bibnamefont {Kolodrubetz}}, \bibinfo {author} {\bibfnamefont
  {Z.}~\bibnamefont {Chen}}, \bibinfo {author} {\bibfnamefont {A.}~\bibnamefont
  {Megrant}}, \bibinfo {author} {\bibfnamefont {R.}~\bibnamefont {Barends}},
  \bibinfo {author} {\bibfnamefont {B.}~\bibnamefont {Campbell}}, \bibinfo
  {author} {\bibfnamefont {B.}~\bibnamefont {Chiaro}}, \bibinfo {author}
  {\bibfnamefont {A.}~\bibnamefont {Dunsworth}}, \bibinfo {author}
  {\bibfnamefont {E.}~\bibnamefont {Jeffrey}}, \bibinfo {author} {\bibfnamefont
  {J.}~\bibnamefont {Kelly}}, \bibinfo {author} {\bibfnamefont
  {J.}~\bibnamefont {Mutus}}, \bibinfo {author} {\bibfnamefont {P.~J.}\
  \bibnamefont {O'Malley}}, \bibinfo {author} {\bibfnamefont {C.}~\bibnamefont
  {Quintana}}, \bibinfo {author} {\bibfnamefont {D.}~\bibnamefont {Sank}},
  \bibinfo {author} {\bibfnamefont {A.}~\bibnamefont {Vainsencher}}, \bibinfo
  {author} {\bibfnamefont {J.}~\bibnamefont {Wenner}}, \bibinfo {author}
  {\bibfnamefont {T.~C.}\ \bibnamefont {White}}, \bibinfo {author}
  {\bibfnamefont {A.}~\bibnamefont {Polkovnikov}}, \ and\ \bibinfo {author}
  {\bibfnamefont {J.~M.}\ \bibnamefont {Martinis}},\ }\href {\doibase
  10.1038/nphys3830} {\bibfield  {journal} {\bibinfo  {journal} {Nature
  Physics}\ }\textbf {\bibinfo {volume} {12}},\ \bibinfo {pages} {1037}
  (\bibinfo {year} {2016})}\BibitemShut {NoStop}%
\bibitem [{\citenamefont {Kaufman}\ \emph {et~al.}(2016)\citenamefont
  {Kaufman}, \citenamefont {{Eric Tai}}, \citenamefont {Lukin}, \citenamefont
  {Rispoli}, \citenamefont {Schittko}, \citenamefont {Preiss},\ and\
  \citenamefont {Greiner}}]{Kaufman2016a}%
  \BibitemOpen
  \bibfield  {author} {\bibinfo {author} {\bibfnamefont {A.~M.}\ \bibnamefont
  {Kaufman}}, \bibinfo {author} {\bibfnamefont {M.}~\bibnamefont {{Eric Tai}}},
  \bibinfo {author} {\bibfnamefont {A.}~\bibnamefont {Lukin}}, \bibinfo
  {author} {\bibfnamefont {M.}~\bibnamefont {Rispoli}}, \bibinfo {author}
  {\bibfnamefont {R.}~\bibnamefont {Schittko}}, \bibinfo {author}
  {\bibfnamefont {P.~M.}\ \bibnamefont {Preiss}}, \ and\ \bibinfo {author}
  {\bibfnamefont {M.}~\bibnamefont {Greiner}},\ }\href {\doibase
  10.1126/science.aaf7894} {\bibfield  {journal} {\bibinfo  {journal}
  {Science}\ }\textbf {\bibinfo {volume} {353}},\ \bibinfo {pages} {790}
  (\bibinfo {year} {2016})}\BibitemShut {NoStop}%
\bibitem [{\citenamefont {Clos}\ \emph {et~al.}(2016)\citenamefont {Clos},
  \citenamefont {Porras}, \citenamefont {Warring},\ and\ \citenamefont
  {Schaetz}}]{Clos2016a}%
  \BibitemOpen
  \bibfield  {author} {\bibinfo {author} {\bibfnamefont {G.}~\bibnamefont
  {Clos}}, \bibinfo {author} {\bibfnamefont {D.}~\bibnamefont {Porras}},
  \bibinfo {author} {\bibfnamefont {U.}~\bibnamefont {Warring}}, \ and\
  \bibinfo {author} {\bibfnamefont {T.}~\bibnamefont {Schaetz}},\ }\href
  {\doibase 10.1103/PhysRevLett.117.170401} {\bibfield  {journal} {\bibinfo
  {journal} {Phys. Rev. Lett.}\ }\textbf {\bibinfo {volume} {117}},\ \bibinfo
  {pages} {170401} (\bibinfo {year} {2016})}\BibitemShut {NoStop}%
\bibitem [{\citenamefont {Deutsch}(1991)}]{Deutsch1991}%
  \BibitemOpen
  \bibfield  {author} {\bibinfo {author} {\bibfnamefont {J.~M.}\ \bibnamefont
  {Deutsch}},\ }\href {\doibase 10.1103/PhysRevA.43.2046} {\bibfield  {journal}
  {\bibinfo  {journal} {Physical Review A}\ }\textbf {\bibinfo {volume} {43}},\
  \bibinfo {pages} {2046} (\bibinfo {year} {1991})}\BibitemShut {NoStop}%
\bibitem [{\citenamefont {Srednicki}(1996)}]{Srednicki1996}%
  \BibitemOpen
  \bibfield  {author} {\bibinfo {author} {\bibfnamefont {M.}~\bibnamefont
  {Srednicki}},\ }\href {\doibase 10.1088/0305-4470/29/4/003} {\bibfield
  {journal} {\bibinfo  {journal} {Journal of Physics A: Mathematical and
  General}\ }\textbf {\bibinfo {volume} {29}},\ \bibinfo {pages} {75} (\bibinfo
  {year} {1996})},\ \Eprint {http://arxiv.org/abs/9511001} {9511001 [chao-dyn]}
  \BibitemShut {NoStop}%
\bibitem [{\citenamefont {Kim}\ \emph {et~al.}(2014)\citenamefont {Kim},
  \citenamefont {Ikeda},\ and\ \citenamefont {Huse}}]{Kim2014}%
  \BibitemOpen
  \bibfield  {author} {\bibinfo {author} {\bibfnamefont {H.}~\bibnamefont
  {Kim}}, \bibinfo {author} {\bibfnamefont {T.~N.}\ \bibnamefont {Ikeda}}, \
  and\ \bibinfo {author} {\bibfnamefont {D.~A.}\ \bibnamefont {Huse}},\ }\href
  {https://arxiv.org/pdf/1408.0535.pdf} {\bibfield  {journal} {\bibinfo
  {journal} {Physical Review E - Statistical, Nonlinear, and Soft Matter
  Physics}\ }\textbf {\bibinfo {volume} {90}} (\bibinfo {year}
  {2014})}\BibitemShut {NoStop}%
\bibitem [{\citenamefont {Mondaini}\ \emph {et~al.}(2016)\citenamefont
  {Mondaini}, \citenamefont {Fratus}, \citenamefont {Srednicki},\ and\
  \citenamefont {Rigol}}]{Mondaini2016}%
  \BibitemOpen
  \bibfield  {author} {\bibinfo {author} {\bibfnamefont {R.}~\bibnamefont
  {Mondaini}}, \bibinfo {author} {\bibfnamefont {K.~R.}\ \bibnamefont
  {Fratus}}, \bibinfo {author} {\bibfnamefont {M.}~\bibnamefont {Srednicki}}, \
  and\ \bibinfo {author} {\bibfnamefont {M.}~\bibnamefont {Rigol}},\ }\href
  {\doibase 10.1103/PhysRevE.93.032104} {\bibfield  {journal} {\bibinfo
  {journal} {Phys. Rev. E}\ }\textbf {\bibinfo {volume} {93}},\ \bibinfo
  {pages} {032104} (\bibinfo {year} {2016})}\BibitemShut {NoStop}%
\bibitem [{\citenamefont {Mondaini}\ and\ \citenamefont
  {Rigol}(2017)}]{Mondaini2017}%
  \BibitemOpen
  \bibfield  {author} {\bibinfo {author} {\bibfnamefont {R.}~\bibnamefont
  {Mondaini}}\ and\ \bibinfo {author} {\bibfnamefont {M.}~\bibnamefont
  {Rigol}},\ }\href {\doibase 10.1103/PhysRevE.96.012157} {\bibfield  {journal}
  {\bibinfo  {journal} {Phys. Rev. E}\ }\textbf {\bibinfo {volume} {96}},\
  \bibinfo {pages} {012157} (\bibinfo {year} {2017})}\BibitemShut {NoStop}%
\bibitem [{\citenamefont {Srednicki}(1999)}]{Srednicki1999}%
  \BibitemOpen
  \bibfield  {author} {\bibinfo {author} {\bibfnamefont {M.}~\bibnamefont
  {Srednicki}},\ }\href {\doibase 10.1088/0305-4470/32/7/007} {\bibfield
  {journal} {\bibinfo  {journal} {J. Phys. A: Math. Gen.}\ }\textbf {\bibinfo
  {volume} {32}},\ \bibinfo {pages} {1163} (\bibinfo {year}
  {1999})}\BibitemShut {NoStop}%
\bibitem [{\citenamefont {Reimann}(2008)}]{Reimann2008}%
  \BibitemOpen
  \bibfield  {author} {\bibinfo {author} {\bibfnamefont {P.}~\bibnamefont
  {Reimann}},\ }\href {\doibase 10.1103/PhysRevLett.101.190403} {\bibfield
  {journal} {\bibinfo  {journal} {Phys. Rev. Lett.}\ }\textbf {\bibinfo
  {volume} {101}},\ \bibinfo {pages} {190403} (\bibinfo {year}
  {2008})}\BibitemShut {NoStop}%
\bibitem [{\citenamefont {Nation}\ and\ \citenamefont
  {Porras}(2018)}]{Nation2018}%
  \BibitemOpen
  \bibfield  {author} {\bibinfo {author} {\bibfnamefont {C.}~\bibnamefont
  {Nation}}\ and\ \bibinfo {author} {\bibfnamefont {D.}~\bibnamefont
  {Porras}},\ }\href
  {http://iopscience.iop.org/article/10.1088/1367-2630/aae28f} {\bibfield
  {journal} {\bibinfo  {journal} {New J. Phys.}\ }\textbf {\bibinfo {volume}
  {20}},\ \bibinfo {pages} {103003} (\bibinfo {year} {2018})}\BibitemShut
  {NoStop}%
\bibitem [{\citenamefont {Nation}\ and\ \citenamefont
  {Porras}(2019{\natexlab{a}})}]{Nation2019}%
  \BibitemOpen
  \bibfield  {author} {\bibinfo {author} {\bibfnamefont {C.}~\bibnamefont
  {Nation}}\ and\ \bibinfo {author} {\bibfnamefont {D.}~\bibnamefont
  {Porras}},\ }\href {\doibase 10.1103/PhysRevE.99.052139} {\bibfield
  {journal} {\bibinfo  {journal} {Physical Review E}\ }\textbf {\bibinfo
  {volume} {99}},\ \bibinfo {pages} {052139} (\bibinfo {year}
  {2019}{\natexlab{a}})}\BibitemShut {NoStop}%
\bibitem [{\citenamefont {Dabelow}\ and\ \citenamefont
  {Reimann}(2019)}]{Dabelow2019}%
  \BibitemOpen
  \bibfield  {author} {\bibinfo {author} {\bibfnamefont {L.}~\bibnamefont
  {Dabelow}}\ and\ \bibinfo {author} {\bibfnamefont {P.}~\bibnamefont
  {Reimann}},\ }\href {https://arxiv.org/pdf/1903.11881.pdf
  http://arxiv.org/abs/1903.11881} {\  (\bibinfo {year} {2019})},\ \Eprint
  {http://arxiv.org/abs/1903.11881} {arXiv:1903.11881} \BibitemShut {NoStop}%
\bibitem [{\citenamefont {Cassidy}\ \emph {et~al.}(2011)\citenamefont
  {Cassidy}, \citenamefont {Clark},\ and\ \citenamefont {Rigol}}]{Cassidy2011}%
  \BibitemOpen
  \bibfield  {author} {\bibinfo {author} {\bibfnamefont {A.~C.}\ \bibnamefont
  {Cassidy}}, \bibinfo {author} {\bibfnamefont {C.~W.}\ \bibnamefont {Clark}},
  \ and\ \bibinfo {author} {\bibfnamefont {M.}~\bibnamefont {Rigol}},\ }\href
  {\doibase 10.1103/PhysRevLett.106.140405} {\bibfield  {journal} {\bibinfo
  {journal} {Phys. Rev. Lett.}\ }\textbf {\bibinfo {volume} {106}},\ \bibinfo
  {pages} {140405} (\bibinfo {year} {2011})}\BibitemShut {NoStop}%
\bibitem [{\citenamefont {Malabarba}\ \emph {et~al.}(2014)\citenamefont
  {Malabarba}, \citenamefont {Garcia-Pintos}, \citenamefont {Linden},
  \citenamefont {Farrelly},\ and\ \citenamefont {Short}}]{Malabarba2014}%
  \BibitemOpen
  \bibfield  {author} {\bibinfo {author} {\bibfnamefont {A.~S.~L.}\
  \bibnamefont {Malabarba}}, \bibinfo {author} {\bibfnamefont {L.~P.}\
  \bibnamefont {Garcia-Pintos}}, \bibinfo {author} {\bibfnamefont
  {N.}~\bibnamefont {Linden}}, \bibinfo {author} {\bibfnamefont {T.~C.}\
  \bibnamefont {Farrelly}}, \ and\ \bibinfo {author} {\bibfnamefont {A.~J.}\
  \bibnamefont {Short}},\ }\href@noop {} {\bibfield  {journal} {\bibinfo
  {journal} {Physical Review E - Statistical, Nonlinear, and Soft Matter
  Physics}\ }\textbf {\bibinfo {volume} {90}} (\bibinfo {year}
  {2014})}\BibitemShut {NoStop}%
\bibitem [{\citenamefont {Garc\'{\i}a-Pintos}\ \emph
  {et~al.}(2017)\citenamefont {Garc\'{\i}a-Pintos}, \citenamefont {Linden},
  \citenamefont {Malabarba}, \citenamefont {Short},\ and\ \citenamefont
  {Winter}}]{Garcia-Pintos2017}%
  \BibitemOpen
  \bibfield  {author} {\bibinfo {author} {\bibfnamefont {L.~P.}\ \bibnamefont
  {Garc\'{\i}a-Pintos}}, \bibinfo {author} {\bibfnamefont {N.}~\bibnamefont
  {Linden}}, \bibinfo {author} {\bibfnamefont {A.~S.~L.}\ \bibnamefont
  {Malabarba}}, \bibinfo {author} {\bibfnamefont {A.~J.}\ \bibnamefont
  {Short}}, \ and\ \bibinfo {author} {\bibfnamefont {A.}~\bibnamefont
  {Winter}},\ }\href {\doibase 10.1103/PhysRevX.7.031027} {\bibfield  {journal}
  {\bibinfo  {journal} {Phys. Rev. X}\ }\textbf {\bibinfo {volume} {7}},\
  \bibinfo {pages} {031027} (\bibinfo {year} {2017})}\BibitemShut {NoStop}%
\bibitem [{\citenamefont {Reimann}(2016)}]{Reimann2016}%
  \BibitemOpen
  \bibfield  {author} {\bibinfo {author} {\bibfnamefont {P.}~\bibnamefont
  {Reimann}},\ }\href {https://www.nature.com/articles/ncomms10821.pdf}
  {\bibfield  {journal} {\bibinfo  {journal} {Nature Communications}\ }\textbf
  {\bibinfo {volume} {7}},\ \bibinfo {pages} {10821} (\bibinfo {year}
  {2016})}\BibitemShut {NoStop}%
\bibitem [{\citenamefont {Borgonovi}\ \emph {et~al.}(2019)\citenamefont
  {Borgonovi}, \citenamefont {Izrailev},\ and\ \citenamefont
  {Santos}}]{Borgonovi2019}%
  \BibitemOpen
  \bibfield  {author} {\bibinfo {author} {\bibfnamefont {F.}~\bibnamefont
  {Borgonovi}}, \bibinfo {author} {\bibfnamefont {F.~M.}\ \bibnamefont
  {Izrailev}}, \ and\ \bibinfo {author} {\bibfnamefont {L.~F.}\ \bibnamefont
  {Santos}},\ }\href {\doibase 10.1103/PhysRevE.99.010101} {\bibfield
  {journal} {\bibinfo  {journal} {Phys. Rev. E}\ }\textbf {\bibinfo {volume}
  {99}},\ \bibinfo {pages} {010101(R)} (\bibinfo {year} {2019})}\BibitemShut
  {NoStop}%
\bibitem [{\citenamefont {Alhambra}\ \emph {et~al.}(2019)\citenamefont
  {Alhambra}, \citenamefont {Riddell},\ and\ \citenamefont
  {Garc{\'{i}}a-Pintos}}]{Alhambra2019}%
  \BibitemOpen
  \bibfield  {author} {\bibinfo {author} {\bibfnamefont {{\'{A}}.~M.}\
  \bibnamefont {Alhambra}}, \bibinfo {author} {\bibfnamefont {J.}~\bibnamefont
  {Riddell}}, \ and\ \bibinfo {author} {\bibfnamefont {L.~P.}\ \bibnamefont
  {Garc{\'{i}}a-Pintos}},\ }\href {https://arxiv.org/pdf/1906.11280.pdf
  http://arxiv.org/abs/1906.11280} {\  (\bibinfo {year} {2019})},\ \Eprint
  {http://arxiv.org/abs/1906.11280} {arXiv:1906.11280} \BibitemShut {NoStop}%
\bibitem [{\citenamefont {Da\ifmmode~\breve{g}\else \u{g}\fi{}}\ \emph
  {et~al.}(2019)\citenamefont {Da\ifmmode~\breve{g}\else \u{g}\fi{}},
  \citenamefont {Sun},\ and\ \citenamefont {Duan}}]{Dag2019}%
  \BibitemOpen
  \bibfield  {author} {\bibinfo {author} {\bibfnamefont {C.~B.}\ \bibnamefont
  {Da\ifmmode~\breve{g}\else \u{g}\fi{}}}, \bibinfo {author} {\bibfnamefont
  {K.}~\bibnamefont {Sun}}, \ and\ \bibinfo {author} {\bibfnamefont {L.-M.}\
  \bibnamefont {Duan}},\ }\href {\doibase 10.1103/PhysRevLett.123.140602}
  {\bibfield  {journal} {\bibinfo  {journal} {Phys. Rev. Lett.}\ }\textbf
  {\bibinfo {volume} {123}},\ \bibinfo {pages} {140602} (\bibinfo {year}
  {2019})}\BibitemShut {NoStop}%
\bibitem [{\citenamefont {Flambaum}\ and\ \citenamefont
  {Izrailev}(1997)}]{Flambaum1997}%
  \BibitemOpen
  \bibfield  {author} {\bibinfo {author} {\bibfnamefont {V.~V.}\ \bibnamefont
  {Flambaum}}\ and\ \bibinfo {author} {\bibfnamefont {F.~M.}\ \bibnamefont
  {Izrailev}},\ }\href {\doibase 10.1103/PhysRevE.56.5144} {\bibfield
  {journal} {\bibinfo  {journal} {Phys. Rev. E}\ }\textbf {\bibinfo {volume}
  {56}},\ \bibinfo {pages} {5144} (\bibinfo {year} {1997})}\BibitemShut
  {NoStop}%
\bibitem [{\citenamefont {Reimann}(2015)}]{Reimann2015}%
  \BibitemOpen
  \bibfield  {author} {\bibinfo {author} {\bibfnamefont {P.}~\bibnamefont
  {Reimann}},\ }\href@noop {} {\bibfield  {journal} {\bibinfo  {journal} {New
  J. Phys.}\ }\textbf {\bibinfo {volume} {17}},\ \bibinfo {pages} {055025}
  (\bibinfo {year} {2015})}\BibitemShut {NoStop}%
\bibitem [{\citenamefont {Torres-Herrera}\ \emph {et~al.}(2016)\citenamefont
  {Torres-Herrera}, \citenamefont {Karp}, \citenamefont {Tavora},\ and\
  \citenamefont {Santos}}]{Torres-Herrera2016}%
  \BibitemOpen
  \bibfield  {author} {\bibinfo {author} {\bibfnamefont {E.~J.}\ \bibnamefont
  {Torres-Herrera}}, \bibinfo {author} {\bibfnamefont {J.}~\bibnamefont
  {Karp}}, \bibinfo {author} {\bibfnamefont {M.}~\bibnamefont {Tavora}}, \ and\
  \bibinfo {author} {\bibfnamefont {L.~F.}\ \bibnamefont {Santos}},\ }\href
  {\doibase 10.3390/e18100359} {\bibfield  {journal} {\bibinfo  {journal}
  {Entropy}\ }\textbf {\bibinfo {volume} {18}},\ \bibinfo {pages} {359}
  (\bibinfo {year} {2016})},\ \Eprint {http://arxiv.org/abs/1608.06636}
  {arXiv:1608.06636} \BibitemShut {NoStop}%
\bibitem [{\citenamefont {Nickelsen}\ and\ \citenamefont
  {Kastner}(2019)}]{Nickelsen2019}%
  \BibitemOpen
  \bibfield  {author} {\bibinfo {author} {\bibfnamefont {D.}~\bibnamefont
  {Nickelsen}}\ and\ \bibinfo {author} {\bibfnamefont {M.}~\bibnamefont
  {Kastner}},\ }\href {http://arxiv.org/abs/1912.02043} {\bibfield  {journal}
  {\bibinfo  {journal} {arXiv:1912.02043}\ } (\bibinfo {year} {2019})},\
  \Eprint {http://arxiv.org/abs/1912.02043} {arXiv:1912.02043} \BibitemShut
  {NoStop}%
\bibitem [{\citenamefont {Nation}\ and\ \citenamefont
  {Porras}(2020)}]{Nation2020}%
  \BibitemOpen
  \bibfield  {author} {\bibinfo {author} {\bibfnamefont {C.}~\bibnamefont
  {Nation}}\ and\ \bibinfo {author} {\bibfnamefont {D.}~\bibnamefont
  {Porras}},\ }\href {\doibase 10.1103/PhysRevE.102.042115} {\bibfield
  {journal} {\bibinfo  {journal} {Phys. Rev. E}\ }\textbf {\bibinfo {volume}
  {102}},\ \bibinfo {pages} {042115} (\bibinfo {year} {2020})}\BibitemShut
  {NoStop}%
\bibitem [{\citenamefont {Neumann}(2010)}]{Neumann2010}%
  \BibitemOpen
  \bibfield  {author} {\bibinfo {author} {\bibfnamefont {J.~V.}\ \bibnamefont
  {Neumann}},\ }\href {\doibase 10.1140/epjh/e2010-00008-5} {\bibfield
  {journal} {\bibinfo  {journal} {The European Physical Journal H}\ }\textbf
  {\bibinfo {volume} {237}},\ \bibinfo {pages} {41} (\bibinfo {year} {2010})},\
  \Eprint {http://arxiv.org/abs/1003.2133} {arXiv:1003.2133} \BibitemShut
  {NoStop}%
\bibitem [{\citenamefont {Goldstein}\ \emph {et~al.}(2009)\citenamefont
  {Goldstein}, \citenamefont {Lebowitz}, \citenamefont {Mastrodonato},
  \citenamefont {Tumulka},\ and\ \citenamefont {Zangh}}]{Goldstein2009}%
  \BibitemOpen
  \bibfield  {author} {\bibinfo {author} {\bibfnamefont {S.}~\bibnamefont
  {Goldstein}}, \bibinfo {author} {\bibfnamefont {J.~L.}\ \bibnamefont
  {Lebowitz}}, \bibinfo {author} {\bibfnamefont {C.}~\bibnamefont
  {Mastrodonato}}, \bibinfo {author} {\bibfnamefont {R.}~\bibnamefont
  {Tumulka}}, \ and\ \bibinfo {author} {\bibfnamefont {N.}~\bibnamefont
  {Zangh}},\ }\href@noop {} {\bibfield  {journal} {\bibinfo  {journal}
  {arXiv:0911.1724}\ ,\ \bibinfo {pages} {1}} (\bibinfo {year} {2009})},\
  \Eprint {http://arxiv.org/abs/arXiv:0911.1724} {arXiv:arXiv:0911.1724}
  \BibitemShut {NoStop}%
\bibitem [{\citenamefont {Vidmar}\ and\ \citenamefont
  {Rigol}(2016)}]{Vidmar2016a}%
  \BibitemOpen
  \bibfield  {author} {\bibinfo {author} {\bibfnamefont {L.}~\bibnamefont
  {Vidmar}}\ and\ \bibinfo {author} {\bibfnamefont {M.}~\bibnamefont {Rigol}},\
  }\href {https://arxiv.org/pdf/1604.03990.pdf} {\bibfield  {journal} {\bibinfo
   {journal} {Journal of Statistical Mechanics: Theory and Experiment}\
  }\textbf {\bibinfo {volume} {2016}} (\bibinfo {year} {2016})}\BibitemShut
  {NoStop}%
\bibitem [{\citenamefont {Nation}\ and\ \citenamefont
  {Porras}(2019{\natexlab{b}})}]{Nation2019a}%
  \BibitemOpen
  \bibfield  {author} {\bibinfo {author} {\bibfnamefont {C.}~\bibnamefont
  {Nation}}\ and\ \bibinfo {author} {\bibfnamefont {D.}~\bibnamefont
  {Porras}},\ }\href {\doibase 10.22331/q-2019-12-02-207} {\bibfield  {journal}
  {\bibinfo  {journal} {{Quantum}}\ }\textbf {\bibinfo {volume} {3}},\ \bibinfo
  {pages} {207} (\bibinfo {year} {2019}{\natexlab{b}})}\BibitemShut {NoStop}%
\bibitem [{\citenamefont {Borgonovi}\ \emph {et~al.}(2017)\citenamefont
  {Borgonovi}, \citenamefont {Mattiotti},\ and\ \citenamefont
  {Izrailev}}]{Borgonovi2017}%
  \BibitemOpen
  \bibfield  {author} {\bibinfo {author} {\bibfnamefont {F.}~\bibnamefont
  {Borgonovi}}, \bibinfo {author} {\bibfnamefont {F.}~\bibnamefont
  {Mattiotti}}, \ and\ \bibinfo {author} {\bibfnamefont {F.~M.}\ \bibnamefont
  {Izrailev}},\ }\href
  {https://journals.aps.org/pre/pdf/10.1103/PhysRevE.95.042135} {\bibfield
  {journal} {\bibinfo  {journal} {Physical Review E}\ }\textbf {\bibinfo
  {volume} {95}},\ \bibinfo {pages} {42135} (\bibinfo {year}
  {2017})}\BibitemShut {NoStop}%
\bibitem [{\citenamefont {Short}(2011)}]{Short2011}%
  \BibitemOpen
  \bibfield  {author} {\bibinfo {author} {\bibfnamefont {A.~J.}\ \bibnamefont
  {Short}},\ }\href
  {http://stacks.iop.org/1367-2630/13/i=5/a=053009?key=crossref.f3a1affa57475206826bc956b2fa8e9a}
  {\bibfield  {journal} {\bibinfo  {journal} {New Journal of Physics}\ }\textbf
  {\bibinfo {volume} {13}},\ \bibinfo {pages} {053009} (\bibinfo {year}
  {2011})}\BibitemShut {NoStop}%
\bibitem [{Note1()}]{Note1}%
  \BibitemOpen
  \bibinfo {note} {Note that our definition of IPR differs here from that in
  \cite {Clos2016a} and in other works in the field of quantum chaos, where the
  reciprocal quantity is defined as the IPR. Our definition here is more
  consistent with the original notion of participation ratio as the number of
  energy eigenstates or atomic orbitals involved in the initial state (see e.g.
  D J Thouless 1974 Phys. Rep.(section C of Physics Letters) 13
  93142).}\BibitemShut {Stop}%
\bibitem [{\citenamefont {Borgonovi}\ and\ \citenamefont
  {Izrailev}(2019)}]{Borgonovi2019a}%
  \BibitemOpen
  \bibfield  {author} {\bibinfo {author} {\bibfnamefont {F.}~\bibnamefont
  {Borgonovi}}\ and\ \bibinfo {author} {\bibfnamefont {F.~M.}\ \bibnamefont
  {Izrailev}},\ }\href {\doibase 10.1103/PhysRevE.99.012115} {\bibfield
  {journal} {\bibinfo  {journal} {Physical Review E}\ }\textbf {\bibinfo
  {volume} {99}},\ \bibinfo {pages} {12115} (\bibinfo {year}
  {2019})}\BibitemShut {NoStop}%
\bibitem [{\citenamefont {Pietracaprina}\ \emph {et~al.}(2017)\citenamefont
  {Pietracaprina}, \citenamefont {Gogolin},\ and\ \citenamefont
  {Goold}}]{Pietracaprina2017}%
  \BibitemOpen
  \bibfield  {author} {\bibinfo {author} {\bibfnamefont {F.}~\bibnamefont
  {Pietracaprina}}, \bibinfo {author} {\bibfnamefont {C.}~\bibnamefont
  {Gogolin}}, \ and\ \bibinfo {author} {\bibfnamefont {J.}~\bibnamefont
  {Goold}},\ }\href {\doibase 10.1103/PhysRevB.95.125118} {\bibfield  {journal}
  {\bibinfo  {journal} {Physical Review B}\ }\textbf {\bibinfo {volume} {95}}
  (\bibinfo {year} {2017}),\ 10.1103/PhysRevB.95.125118}\BibitemShut {NoStop}%
\bibitem [{\citenamefont {Venuti}\ and\ \citenamefont
  {Liu}(2019)}]{Venuti2019}%
  \BibitemOpen
  \bibfield  {author} {\bibinfo {author} {\bibfnamefont {L.~C.}\ \bibnamefont
  {Venuti}}\ and\ \bibinfo {author} {\bibfnamefont {L.}~\bibnamefont {Liu}},\
  }\href {https://arxiv.org/pdf/1904.02336.pdf http://arxiv.org/abs/1904.02336}
  {\  (\bibinfo {year} {2019})},\ \Eprint {http://arxiv.org/abs/1904.02336}
  {arXiv:1904.02336} \BibitemShut {NoStop}%
\bibitem [{\citenamefont {Goldstein}\ \emph {et~al.}(2006)\citenamefont
  {Goldstein}, \citenamefont {Lebowitz}, \citenamefont {Tumulka},\ and\
  \citenamefont {Zangh\`{\i}}}]{Goldstein2006}%
  \BibitemOpen
  \bibfield  {author} {\bibinfo {author} {\bibfnamefont {S.}~\bibnamefont
  {Goldstein}}, \bibinfo {author} {\bibfnamefont {J.~L.}\ \bibnamefont
  {Lebowitz}}, \bibinfo {author} {\bibfnamefont {R.}~\bibnamefont {Tumulka}}, \
  and\ \bibinfo {author} {\bibfnamefont {N.}~\bibnamefont {Zangh\`{\i}}},\
  }\href {\doibase 10.1103/PhysRevLett.96.050403} {\bibfield  {journal}
  {\bibinfo  {journal} {Phys. Rev. Lett.}\ }\textbf {\bibinfo {volume} {96}},\
  \bibinfo {pages} {050403} (\bibinfo {year} {2006})}\BibitemShut {NoStop}%
\bibitem [{\citenamefont {Reimann}(2007)}]{Reimann2007}%
  \BibitemOpen
  \bibfield  {author} {\bibinfo {author} {\bibfnamefont {P.}~\bibnamefont
  {Reimann}},\ }\href {\doibase 10.1103/PhysRevLett.99.160404} {\bibfield
  {journal} {\bibinfo  {journal} {Physical Review Letters}\ }\textbf {\bibinfo
  {volume} {99}},\ \bibinfo {pages} {160404} (\bibinfo {year} {2007})},\
  \Eprint {http://arxiv.org/abs/0710.4214} {arXiv:0710.4214} \BibitemShut
  {NoStop}%
\bibitem [{\citenamefont {Popescu}\ \emph {et~al.}(2006)\citenamefont
  {Popescu}, \citenamefont {Short},\ and\ \citenamefont
  {Winter}}]{Popescu2006}%
  \BibitemOpen
  \bibfield  {author} {\bibinfo {author} {\bibfnamefont {S.}~\bibnamefont
  {Popescu}}, \bibinfo {author} {\bibfnamefont {A.~J.}\ \bibnamefont {Short}},
  \ and\ \bibinfo {author} {\bibfnamefont {A.}~\bibnamefont {Winter}},\ }\href
  {\doibase 10.1038/nphys444} {\bibfield  {journal} {\bibinfo  {journal}
  {Nature Physics}\ }\textbf {\bibinfo {volume} {2}},\ \bibinfo {pages} {754}
  (\bibinfo {year} {2006})}\BibitemShut {NoStop}%
\bibitem [{\citenamefont {Kubo}(1966)}]{Kubo1966}%
  \BibitemOpen
  \bibfield  {author} {\bibinfo {author} {\bibfnamefont {R.}~\bibnamefont
  {Kubo}},\ }\href {\doibase 10.1088/0034-4885/29/1/306} {\bibfield  {journal}
  {\bibinfo  {journal} {Rep. Prog. Phys.}\ }\textbf {\bibinfo {volume} {29}},\
  \bibinfo {pages} {255} (\bibinfo {year} {1966})}\BibitemShut {NoStop}%
\bibitem [{\citenamefont {Breuer}\ and\ \citenamefont
  {Petruccione}(2002)}]{Breuer2002}%
  \BibitemOpen
  \bibfield  {author} {\bibinfo {author} {\bibfnamefont {H.~P.}\ \bibnamefont
  {Breuer}}\ and\ \bibinfo {author} {\bibfnamefont {F.}~\bibnamefont
  {Petruccione}},\ }\href@noop {} {\emph {\bibinfo {title} {The theory of open
  quantum systems}}}\ (\bibinfo  {publisher} {Oxford University Press},\
  \bibinfo {address} {Great Clarendon Street},\ \bibinfo {year}
  {2002})\BibitemShut {NoStop}%
\bibitem [{\citenamefont {Maldacena}\ \emph {et~al.}(2016)\citenamefont
  {Maldacena}, \citenamefont {Shenker},\ and\ \citenamefont
  {Stanford}}]{Maldacena2016}%
  \BibitemOpen
  \bibfield  {author} {\bibinfo {author} {\bibfnamefont {J.}~\bibnamefont
  {Maldacena}}, \bibinfo {author} {\bibfnamefont {S.~H.}\ \bibnamefont
  {Shenker}}, \ and\ \bibinfo {author} {\bibfnamefont {D.}~\bibnamefont
  {Stanford}},\ }\href {\doibase 10.1007/JHEP08(2016)106} {\bibfield  {journal}
  {\bibinfo  {journal} {Journal of High Energy Physics}\ }\textbf {\bibinfo
  {volume} {2016}},\ \bibinfo {pages} {106} (\bibinfo {year}
  {2016})}\BibitemShut {NoStop}%
\bibitem [{\citenamefont {Hashimoto}\ \emph {et~al.}(2017)\citenamefont
  {Hashimoto}, \citenamefont {Murata},\ and\ \citenamefont
  {Yoshii}}]{Hashimoto2017}%
  \BibitemOpen
  \bibfield  {author} {\bibinfo {author} {\bibfnamefont {K.}~\bibnamefont
  {Hashimoto}}, \bibinfo {author} {\bibfnamefont {K.}~\bibnamefont {Murata}}, \
  and\ \bibinfo {author} {\bibfnamefont {R.}~\bibnamefont {Yoshii}},\ }\href
  {https://arxiv.org/pdf/1703.09435.pdf} {\bibfield  {journal} {\bibinfo
  {journal} {Journal of High Energy Physics}\ }\textbf {\bibinfo {volume}
  {2017}} (\bibinfo {year} {2017})}\BibitemShut {NoStop}%
\bibitem [{\citenamefont {Yan}\ \emph {et~al.}(2019)\citenamefont {Yan},
  \citenamefont {Cincio},\ and\ \citenamefont {Zurek}}]{Yan2019}%
  \BibitemOpen
  \bibfield  {author} {\bibinfo {author} {\bibfnamefont {B.}~\bibnamefont
  {Yan}}, \bibinfo {author} {\bibfnamefont {L.}~\bibnamefont {Cincio}}, \ and\
  \bibinfo {author} {\bibfnamefont {W.~H.}\ \bibnamefont {Zurek}},\ }\href
  {https://arxiv.org/pdf/1903.02651.pdf http://arxiv.org/abs/1903.02651} {\
  (\bibinfo {year} {2019})},\ \Eprint {http://arxiv.org/abs/1903.02651}
  {arXiv:1903.02651} \BibitemShut {NoStop}%
\bibitem [{\citenamefont {Swingle}(2018)}]{Swingle2018}%
  \BibitemOpen
  \bibfield  {author} {\bibinfo {author} {\bibfnamefont {B.}~\bibnamefont
  {Swingle}},\ }\href {\doibase 10.1038/s41567-018-0295-5} {\bibfield
  {journal} {\bibinfo  {journal} {Nature Physics}\ }\textbf {\bibinfo {volume}
  {14}},\ \bibinfo {pages} {988} (\bibinfo {year} {2018})}\BibitemShut
  {NoStop}%
\bibitem [{\citenamefont {Iyoda}\ and\ \citenamefont
  {Sagawa}(2018)}]{Iyoda2017}%
  \BibitemOpen
  \bibfield  {author} {\bibinfo {author} {\bibfnamefont {E.}~\bibnamefont
  {Iyoda}}\ and\ \bibinfo {author} {\bibfnamefont {T.}~\bibnamefont {Sagawa}},\
  }\href {\doibase 10.1103/PhysRevA.97.042330} {\bibfield  {journal} {\bibinfo
  {journal} {Phys. Rev. A}\ }\textbf {\bibinfo {volume} {97}},\ \bibinfo
  {pages} {042330} (\bibinfo {year} {2018})}\BibitemShut {NoStop}%
\bibitem [{\citenamefont {Huang}\ \emph {et~al.}(2019)\citenamefont {Huang},
  \citenamefont {Brand\~ao},\ and\ \citenamefont {Zhang}}]{Huang2017}%
  \BibitemOpen
  \bibfield  {author} {\bibinfo {author} {\bibfnamefont {Y.}~\bibnamefont
  {Huang}}, \bibinfo {author} {\bibfnamefont {F.~G. S.~L.}\ \bibnamefont
  {Brand\~ao}}, \ and\ \bibinfo {author} {\bibfnamefont {Y.-L.}\ \bibnamefont
  {Zhang}},\ }\href {\doibase 10.1103/PhysRevLett.123.010601} {\bibfield
  {journal} {\bibinfo  {journal} {Phys. Rev. Lett.}\ }\textbf {\bibinfo
  {volume} {123}},\ \bibinfo {pages} {010601} (\bibinfo {year}
  {2019})}\BibitemShut {NoStop}%
\bibitem [{\citenamefont {Foini}\ and\ \citenamefont
  {Kurchan}(2019)}]{Foini2019}%
  \BibitemOpen
  \bibfield  {author} {\bibinfo {author} {\bibfnamefont {L.}~\bibnamefont
  {Foini}}\ and\ \bibinfo {author} {\bibfnamefont {J.}~\bibnamefont
  {Kurchan}},\ }\href {\doibase 10.1103/PhysRevE.99.042139} {\bibfield
  {journal} {\bibinfo  {journal} {Phys. Rev. E}\ }\textbf {\bibinfo {volume}
  {99}},\ \bibinfo {pages} {042139} (\bibinfo {year} {2019})}\BibitemShut
  {NoStop}%
\bibitem [{\citenamefont {Murthy}\ and\ \citenamefont
  {Srednicki}(2019)}]{Murthy2019}%
  \BibitemOpen
  \bibfield  {author} {\bibinfo {author} {\bibfnamefont {C.}~\bibnamefont
  {Murthy}}\ and\ \bibinfo {author} {\bibfnamefont {M.}~\bibnamefont
  {Srednicki}},\ }\href {\doibase 10.1103/PhysRevLett.123.230606} {\bibfield
  {journal} {\bibinfo  {journal} {Phys. Rev. Lett.}\ }\textbf {\bibinfo
  {volume} {123}},\ \bibinfo {pages} {230606} (\bibinfo {year}
  {2019})}\BibitemShut {NoStop}%
\bibitem [{\citenamefont {Roberts}\ and\ \citenamefont
  {Yoshida}(2017)}]{Roberts2017}%
  \BibitemOpen
  \bibfield  {author} {\bibinfo {author} {\bibfnamefont {D.~A.}\ \bibnamefont
  {Roberts}}\ and\ \bibinfo {author} {\bibfnamefont {B.}~\bibnamefont
  {Yoshida}},\ }\href {\doibase 10.1007/JHEP04(2017)121} {\bibfield  {journal}
  {\bibinfo  {journal} {Journal of High Energy Physics}\ }\textbf {\bibinfo
  {volume} {2017}},\ \bibinfo {pages} {121} (\bibinfo {year} {2017})},\ \Eprint
  {http://arxiv.org/abs/1610.04903} {arXiv:1610.04903} \BibitemShut {NoStop}%
\bibitem [{\citenamefont {Da\ifmmode~\breve{g}\else \u{g}\fi{}}\ \emph
  {et~al.}(2020)\citenamefont {Da\ifmmode~\breve{g}\else \u{g}\fi{}},
  \citenamefont {Duan},\ and\ \citenamefont {Sun}}]{Dag2020}%
  \BibitemOpen
  \bibfield  {author} {\bibinfo {author} {\bibfnamefont {C.~B.}\ \bibnamefont
  {Da\ifmmode~\breve{g}\else \u{g}\fi{}}}, \bibinfo {author} {\bibfnamefont
  {L.-M.}\ \bibnamefont {Duan}}, \ and\ \bibinfo {author} {\bibfnamefont
  {K.}~\bibnamefont {Sun}},\ }\href {\doibase 10.1103/PhysRevB.101.104415}
  {\bibfield  {journal} {\bibinfo  {journal} {Phys. Rev. B}\ }\textbf {\bibinfo
  {volume} {101}},\ \bibinfo {pages} {104415} (\bibinfo {year}
  {2020})}\BibitemShut {NoStop}%
\bibitem [{\citenamefont {Chan}\ \emph {et~al.}(2019)\citenamefont {Chan},
  \citenamefont {{De Luca}},\ and\ \citenamefont {Chalker}}]{Chan2019}%
  \BibitemOpen
  \bibfield  {author} {\bibinfo {author} {\bibfnamefont {A.}~\bibnamefont
  {Chan}}, \bibinfo {author} {\bibfnamefont {A.}~\bibnamefont {{De Luca}}}, \
  and\ \bibinfo {author} {\bibfnamefont {J.~T.}\ \bibnamefont {Chalker}},\
  }\href {\doibase 10.1103/PhysRevLett.122.220601} {\bibfield  {journal}
  {\bibinfo  {journal} {Physical Review Letters}\ }\textbf {\bibinfo {volume}
  {122}},\ \bibinfo {pages} {1} (\bibinfo {year} {2019})},\ \Eprint
  {http://arxiv.org/abs/1810.11014} {arXiv:1810.11014} \BibitemShut {NoStop}%
\bibitem [{\citenamefont {Schiulaz}\ \emph {et~al.}(2019)\citenamefont
  {Schiulaz}, \citenamefont {Torres-Herrera}, \citenamefont
  {P{\'{e}}rez-Bernal},\ and\ \citenamefont {Santos}}]{Schiulaz2019a}%
  \BibitemOpen
  \bibfield  {author} {\bibinfo {author} {\bibfnamefont {M.}~\bibnamefont
  {Schiulaz}}, \bibinfo {author} {\bibfnamefont {E.~J.}\ \bibnamefont
  {Torres-Herrera}}, \bibinfo {author} {\bibfnamefont {F.}~\bibnamefont
  {P{\'{e}}rez-Bernal}}, \ and\ \bibinfo {author} {\bibfnamefont {L.~F.}\
  \bibnamefont {Santos}},\ }\href {https://arxiv.org/pdf/1906.11856.pdf
  http://arxiv.org/abs/1906.11856} {\  (\bibinfo {year} {2019})},\ \Eprint
  {http://arxiv.org/abs/1906.11856} {arXiv:1906.11856} \BibitemShut {NoStop}%
\bibitem [{\citenamefont {Weisskopf}\ and\ \citenamefont
  {Wigner}(1930)}]{Weisskopf1930}%
  \BibitemOpen
  \bibfield  {author} {\bibinfo {author} {\bibfnamefont {V.}~\bibnamefont
  {Weisskopf}}\ and\ \bibinfo {author} {\bibfnamefont {E.}~\bibnamefont
  {Wigner}},\ }\href {\doibase 10.1007/BF01336768} {\bibfield  {journal}
  {\bibinfo  {journal} {Zeitschrift f{\"u}r Physik}\ }\textbf {\bibinfo
  {volume} {63}},\ \bibinfo {pages} {54} (\bibinfo {year} {1930})}\BibitemShut
  {NoStop}%
\bibitem [{\citenamefont {Cohen-Tannoudji}\ \emph {et~al.}(1998)\citenamefont
  {Cohen-Tannoudji}, \citenamefont {Dupont-Roc},\ and\ \citenamefont
  {Grynberg}}]{Cohen-Tannoudji}%
  \BibitemOpen
  \bibfield  {author} {\bibinfo {author} {\bibfnamefont {C.}~\bibnamefont
  {Cohen-Tannoudji}}, \bibinfo {author} {\bibfnamefont {J.}~\bibnamefont
  {Dupont-Roc}}, \ and\ \bibinfo {author} {\bibfnamefont {G.}~\bibnamefont
  {Grynberg}},\ }\href {\doibase 10.1063/1.2809840} {\bibfield  {journal}
  {\bibinfo  {journal} {Atom-Photon Interactions: Basic Processes and
  Applications, by Claude Cohen-Tannoudji, Jacques Dupont-Roc, Gilbert
  Grynberg, pp. 678. ISBN 0-471-29336-9. Wiley-VCH , March 1998.}\ }\textbf
  {\bibinfo {volume} {-1}} (\bibinfo {year} {1998}),\
  10.1063/1.2809840}\BibitemShut {NoStop}%
\end{thebibliography}
